\documentclass[12pt, letterpaper]{article}

\usepackage{amsmath,amssymb}%Needed for multline
\usepackage{cite}%Needed for long citations
\usepackage{fancyhdr}%Needed for title page headings
\usepackage[top=1in, bottom=1in, left=1in, right=1in]{geometry}%Needed for margins
\usepackage{graphicx}%Needed for includegraphics
\usepackage{hyperref}%Needed for hyper references

\numberwithin{equation}{section}%Needed for equation numbers to follow sections
\date{}%Needed to remove the date

\begin{document}
\title{{\rm\footnotesize \qquad \qquad \qquad \qquad \qquad \ \qquad \qquad \qquad \ \ \ \ \ \                      RUNHETC-2024-31
}\vskip.5in    Hilbert Bundles and Holographic Space-time: the Hydrodynamic Approach to Gravity}
\author{Tom Banks\\
NHETC and Department of Physics \\
Rutgers University, Piscataway, NJ 08854-8019\\
E-mail: \href{mailto:tibanks@ucsc.edu}{tibanks@ucsc.edu}
\\
\\
%Willy Fischler\\
%Department of Physics and Texas Cosmology Center\\
%Weinberg Institute, Center for Theory\\
%University of Texas, Austin, TX 78712\\
%E-mail: \href{mailto:fischler@physics.utexas.edu}{fischler@physics.utexas.edu}
}

\maketitle
\thispagestyle{fancy} %Needed to remove the page number from the title page

\begin{abstract} \normalsize \noindent  Results of Jacobson, Carlip and Solodukhin, from the 1990s, suggest a hydrodynamic approach to quantum gravity in which a classical solution of Einstein's equations determines the density matrices of subsystems associated with causal diamonds in the "empty diamond" state of a corresponding quantum system.  The subsystem operator algebras are finite dimensional and correspond to a UV cutoff $1 + 1$ dimensional field theory of fermions living on a "stretched horizon" near each diamond's holographic screen. The fields can be thought of as fluctuations of solutions of the screen's Dirac operator around that of the background geometry, expanded up to a maximal Dirac eigenvalue determined by the Carlip-Solodukhin relation between area and central charge.  This cutoff renders the screen geometry "fuzzy".  Quantum dynamics is defined in a Hilbert bundle over the space of time-like geodesics on the background geometry.  A nesting of diamonds along a given geodesic defines a series of unitary embedding maps of diamond Hilbert spaces into each other, analogous to half sided modular flows in quantum field theory.  These can be extended into a consistent set of unitary maps of each fiber Hilbert space into itself by a quantum version of the principle of relativity.  According to the QPR, the largest diamond in the overlap between any two diamonds is identified with a tensor factor in each individual diamond Hilbert space, and must have the a density matrix with the same entanglement spectrum no matter which fiber dynamics is used to compute it.  This brief review summarizes how these ideas play out in a variety of contexts in different dimensions.\end{abstract}

%%%%%%%%%%%%%%%%%%%%%%%%%%%%%%%%%%%%%%%%%%%%%%%%%%%%%%%%%%%%%%%%%%%%%%%%%%%%%%%%%%%%%%%%%%%%%%%%%%%%%%%%%%%%%%%%%%%%%%%%%%%%%%%%%%%%%%%%%%%%%%%%%%%%

\newpage
\tableofcontents
\vspace{1cm}

%%%%%%%%%%%%%%%%%%%%%%%%%%%%%%%%%%%%%%%%%%%%%%%%%%%%%%%%%%%%%%%%%%%%%%%%%%%%%%%%%%%%%%%%%%%%%%%%%%%%%%%%%%%%%%%%%%%%%%%%%%%%%%%%%%%%%%%%%%%%%%%%%%%%
\vfill\eject
\section{Introduction}

The basic object in experimental physics is described theoretically as a detector whose center of mass follows a time-like trajectory through space-time. Our notions of what the words in the previous sentence meant have been modified substantially since the invention of Newtonian classical mechanics.  Classical space-time is dynamical and is modified by the matter moving through it.  Quantum mechanics teaches us that detectors must be large systems, with many microscopic degrees of freedom, {\it and as many macroscopic pointer variables as there are microscopic q-bits in the system the detector intends to probe}.  Einstein argued that the maximal system a detector could probe in proper time $\tau$ had to fit into the {\it causal diamond} of that segment of its trajectory.

Attempts to construct a theory of quantum gravity have revealed a basic conceptual flaw in the first sentence of the previous paragraph.  There is considerable evidence\cite{evidence} that the entropy of any quantum system inside the causal diamond is bounded by $\frac{A_{\diamond}}{4G_N}$, where $A_{\diamond}$ is the $d - 2$ volume of the maximum volume space-like surface in a null foliation of the diamond boundary (henceforth called the "area of the diamond's holographic screen").  Furthermore, most of the quantum states contributing to that entropy come from a black hole completely filling the diamond.  Black holes have many quantum states but very few macroscopic subsystems that can serve as pointers, because they scramble information so rapidly\cite{lshpss}.  Thus, the idea of a detector inside a diamond being the appropriate tool on which to hang a theory of the quantum mechanics of the diamond appears to be a non-starter.

Fortunately, the area law itself points a way out of this difficulty.  The key paper demonstrating this was\cite{ted95}, which showed that Einstein's equations, with an unknown, covariantly conserved, stress tensor, were the hydrodynamic equations (local version of thermodynamics) of this law.  The cosmological constant (c.c.) does not contribute to these hydrodynamic equations, which suggests that it should not be interpreted as "the expectation value of the stress tensor in the lowest energy state of the theory", contrary to the standard interpretation in quantum field theory (QFT).  We will describe the proper interpretation of the c.c. in Section 4 below.

Indeed, the statement above that most of the states in a causal diamond belong to a diamond filling black hole\footnote{In the cosmological context, a diamond filling black hole is unstable.  The phrase is shorthand for the mixed $p = \pm \rho$ flat FRW cosmology, or for the Novikov-Zeldovich\cite{nz} black hole that grows with the horizon of a less stiff FRW cosmology. It's only in inhomogeneous cosmologies that meta-stable black holes of fixed size can exist.}, shows that "QFT in curved space-time" is always a very bad approximation to a real theory of quantum gravity (QG) in any causal diamond.
Cohen, Kaplan and Nelson\cite{ckn} showed that throwing away the QFT states that would make diamond filling black holes does not effect the agreement between QFT and any experiment.  The resulting QFT entropy grows like a subleading power of the diamond area and cannot account for the BHJFSB\cite{evidence} area law.  Despite this, we'll see that QFT can offer us some clues about the nature of QG, which we'll discuss in Section 2.  Indeed, it was by pondering the universal area law entanglement of diamond subalgebras in "empty diamond" states of (cut-off) QFT\cite{sorkinetal}, that Jacobson was led to find a connection between a universal area law for diamond entropy, and Einstein's equations.

A clue to the correct model for the QM of diamonds was found by examining fluctuations around a black hole solution of Einstein's equations, restricted to a small region (the inner stretched horizon) near the black hole horizon.  This was done by Carlip\cite{carlip} and, independently by Solodukhin\cite{solo} in 1998.  In 2021 Zurek and the present author\cite{BZ} realized that the CS argument worked for any causal diamond, in the immediate vicinity of its holographic screen.  The argument is most easily understood in Solodukhin's language.  He shows that the near horizon action for small fluctuations of the transverse spatial conformal factor is that of a massless two dimensional scalar field on the stretched horizon interval, with left moving stress tensor
\begin{equation} T_{zz} = (\partial_z \phi)^2  +  Q   \partial_z^2 \phi  . \end{equation}

This is well known\cite{bpz} to be the "hydrodynamic" stress tensor, whose correlators in the ground state reproduce those of the stress tensor of any conformal field theory (CFT) with central charge 
\begin{equation} c =  3Q^2 \end{equation} in the classical approximation.   One finds that the central charge is proportional to the area of the holographic screen in whatever microscopic units one has used to zoom in on the near horizon limit.  There's a corresponding multiplicative ambiguity in the entropy, but the only natural choice is the Planck scale.  
Carlip and Solodukhin assume that the density matrix is indeed that of such a CFT.

One can then use Cardy's asymptotic formula for the density of states to compute the entropy, and finds consistency with Jacobson's assumption that the entropy is $\frac{A_{\diamond}}{4G_N}$.  CS did this calculation for all black holes.  The observation of\cite{BZ} generalizes it to all causal diamonds, closing the circle and deriving Jacobson's fundamental assumption.  The area law can also be derived by Euclidean path integral methods for many cases\cite{gh1}\cite{bdf}.  In the work of Carlip and Solodukhin, the Cardy formula was evaluated at a fixed Virasoro level, corresponding to a classical solution of the scalar field equations.  In\cite{BZ} we showed that it followed from saddle point approximation of the spectral integral for large central charge.  In either case, it's obvious that the argument holds for cut off CFTs, as long as the cut off is just above the dominant saddle point.  

The authors \cite{BZ} pointed out that the CS analysis implies the fluctuation relation
\begin{equation} \langle (K - \langle K \rangle)^2 \rangle = \langle K \rangle . \end{equation}
This relation had been derived by two independent methods\cite{VZ2}\cite{deBoer} for boundary anchored Ryu-Takayanagi (RT) diamonds in anti-de Sitter (AdS) space.  The calculation of the modular fluctuation in\cite{deBoer} is interesting because it uses the AdS/CFT duality and a general result of Perlmutter\cite{perl} for CFT.  The RT diamond is dual to a boundary causal diamond in Minkowski space and its modular Hamiltonian is dual to that of the CFT\cite{BW}\cite{CHM}.  In general one has
\begin{equation} \langle (K - \langle K \rangle)^2 \rangle = a \langle K \rangle , \end{equation} where $a$ is a model dependent coefficient in the two point function of the stress tensor.  For all CFTs with an Einstein-Hilbert (EH) dual $a \approx 1$, where the approximation means that higher curvature terms in the gravitational action are neglected.  Of course, the CS result is derived under the same assumption.  

In QFT both sides of the equation are state independent in a given model but UV divergent, as long as the states are created by fields acting on the vacuum at points outside the diamond and space-like separated from it.  Their ratio is always finite, but depends on the regulator scheme.  There is a class of regulators, in which one cuts off the spectrum of modes transverse to the horizon in some way, which gives $a = 1$ for scalar and spinor fields.  The conformal regulator of\cite{CHM} instead gives the results of\cite{perl}.   One can also conclude that $a = 1$ in QG for de Sitter (dS) space using a gravitational replica trick\cite{tbpdreplica}.  

One is thus led to the following conjecture to replace the detector centric picture of space-time physics described in the first paragraph: a solution of Einstein's equations defines the hydrodynamics of a hypothetical quantum system, which must somehow assign a density matrix $\frac{e^{- L_0 (A_{\diamond})}}{{\rm Tr}\ e^{- L_0 (A_{\diamond})}} $, to each causal diamond in the space-time.  $L_0$ is the Virasoro generator of a $1 + 1$ dimensional CFT on an interval\footnote{The CFT will actually have a UV cutoff, which we discuss below.}.  Its central charge is given by $c = \propto \frac{A_{\diamond}}{2G_N} $.  These density matrices describe the "empty diamond" state in the background geometry.  For non-negative c.c. this will always be the maximal entropy state in the diamond.  For negative c.c. this description is only valid until the proper time in the diamond is of order the AdS radius.  At larger scales a tensor network description becomes appropriate, as we will discuss below.  

Dynamics in the hypothetical quantum theory is defined in a Hilbert bundle over the space of time-like geodesics in the hydrodynamic background. This space is non-compact and includes null geodesics as limit points.  One must also supply a nesting of proper time intervals, with Planck scale separation along each geodesic.  
\begin{figure}[h]
\begin{center}
\includegraphics[width=01\linewidth]{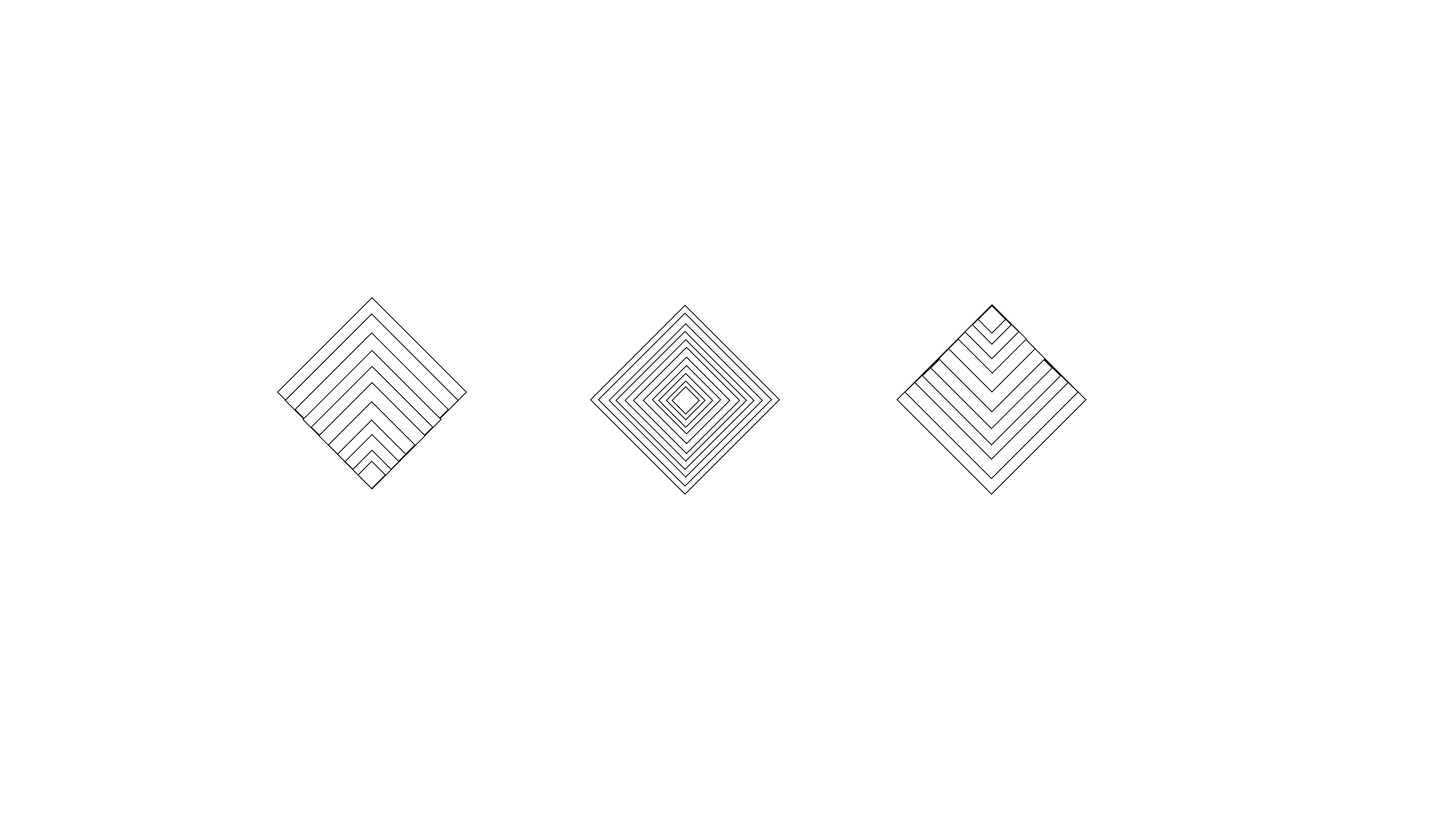}

\caption{Future directed, time symmetric, and past directed nested coverings of a causal diamond.} 
\label{fig:nestedcoversofadiamond}
\end{center}
\end{figure}

Along each geodesic, there is then a natural time dependent evolution operator, morally related (in the case of time-asymmetric nesting of proper time intervals in the diamond) to the half sided modular flow of QFT, within causal diamonds. This is given by the unitary embedding
$ e^{i L_0 (t)} e^{- i L_0 (t + L_P)}$, where $L_0 (\tau)$ is the Virasoro generator of the C-S CFT of the diamond at proper time $\tau$ from the origin of time.  It evolves the system in the wedge between the two nested diamonds and embeds the Hilbert space of the smaller diamond into that of the larger one. This must be supplemented, at each instant, with a prescription for describing evolution on the complementary tensor factor of the Hilbert space\footnote{For non-positive c.c., we must impose a finite area cutoff on diamonds to define the complementary tensor factor, and then carefully take the limit as the cutoff goes to infinity.  For negative c.c. this is the process of holographic renormalization of tensor networks, while for vanishing c.c. the mathematical formalism and a description of the asymptotic Hilbert space still does not, IMHO, exist.}. This describes the universe outside the diamond in some set of coordinates compatible with those defined by modular flow within the diamond.  It is conjectured that this is supplied by a connection on the Hilbert bundle of the hydrodynamical space-time, which is described in words by saying that the largest diamond in the geometric overlap of diamonds along two different geodesics, must correspond to an isomorphic tensor factor in each diamond Hilbert space, whose density matrix has identical entanglement spectrum, when computed in either of the two fibers of the bundle.  This must be true not only for the "empty diamond" state, but also for states corresponding to localized excitations inside diamonds.  When such localized states are described as constraints on the boundary q-bits, it's easy to see pictorially that this principle, called {\it The Quantum Principle of Relativity} (QPR) leads to a description of localized states in terms of trajectories inside diamonds.   We will deal with this in Section 2b.  However, a formulation of the QPR in terms of a convenient set of equations that can be imposed on the holographic variables has not yet been found.  This remains the biggest open problem in the approach to QG advocated here.  

The QPR also describes the relationship between descriptions of the Hilbert bundle that use different nested proper time intervals along time-like trajectories.  For most geometries it's sufficient to restrict attention to timelike geodesics, including the limit points of where some or all of the tangent vectors become null.  Then, for a general space-time we have two natural nestings, starting from either the past or future tips of maximally extended geodesics.  For space-times with a time reflection symmetry we also have a natural nesting expanding out from some surface invariant under the reflection.  This is useful for discussing a possible "scattering theory" interpretation of the observables.  In all cases, gauge invariant observables appear, from the point of view of the hydrodynamic approximation to be functions on the space of time-like geodesics, or equivalently on the asymptotic boundaries which they approach.  For non-positive cosmological constant this hypothesis seems to be validated by perturbative string theory and AdS/CFT.  For positive c.c. it was hypothesized in the dS/CFT approach\cite{dSCFT}, but this seems to be at odds with the finite entropy of dS space\cite{sussetal}.  Earlier\cite{tbwf99} Fischler and the present author had hypothesized that boundary S matrix like observables in dS space made sense as finite time amplitudes, but would eventually fade into equilibrium\cite{tbdsobs}.  In interesting recent series of papers\cite{cem} suggests that even in an asymptotically dS cosmology with infinite entropy, the proper observables must be integrated over the asymptotic boundary in a diffeomorphism invariant manner.  The gauge invariant correlation functions depend on labels of the irreps of the dS group but do not transform under that group.  It is not clear what the quantum mechanical interpretation of these amplitudes is, although a Hilbert space structure appears rather naturally in the formalism.  

As noted by Solodukhin\cite{solo}, while the classical fluctuations of the transverse conformal factor are "taken into account" by the $1 + 1$ CFT prescription, one should also have a way of encoding fluctuations in the rest of the transverse geometry.  In\cite{hilbertbundles} we proposed to do this by incorporating the formalism of Holographic Space Time\cite{hst} (HST).   HST was also motivated by the idea that the entropy in a causal diamond is finite and the fact that finite dimensional Hilbert spaces can always be constructed from constrained fermions.  Connes' observation that Riemannian geometry is entirely encoded in the Dirac operator on a Riemannian manifold\cite{connes} can be combined with the ideas of CS, by constructing the $1 + 1$ dimensional CFT out of fermion fields that are labeled by eigenspinors of the Dirac operator on the transverse geometry.  One can think of the resulting "fluctuating transverse Dirac operator" as describing the fluctuating geometry in the basis provided by the eigenspectrum of the background Dirac operator.  The CS restriction on the central charge of the CFT becomes a UV cutoff on the eigenvalue of the transverse Dirac operator, which provides a "fuzzification" of the manifold that does not depend on a Kahler structure\cite{tbjk}.  When this cutoff becomes so stringent that we only have a single $1 + 1$ dimensional fermion field, the $1 + 1$ dimensional cutoff on the spectrum of that field tells us what the minimal size of causal diamond is for which the semi-classical ideas of CS make sense.  From a purely theoretical point of view, we have to keep enough modes for Cardy's formula to be valid for a single fermion, but apart from that the value of this cutoff can probably only be determined by hypothetical futuristic experiments. 
We'll argue below that the fast scrambling behavior of horizons suggests that the actual CFT is a particular four fermion deformation of free field theory.  

In the next few sections we'll outline why QFT is a bad approximation to a theory of QG obeying these principles and why that doesn't contradict the experimental success of QFT.  We'll explain how the Hilbert bundle formulation does bear a close resemblance to Algebraic QFT (AQFT), with the principle difference being that local operator algebras are finite dimensional and global invariance principles are replaced by the QPR.  We'll outline how global symmetries may arise from the QPR as asymptotic symmetries of asymptotically flat or AdS spaces.  The most important concept will be the realization of local physics in terms of constraints on the states of holographic variables.  This has the advantage of producing a subset of states that behave like the scattering states of (jets of) particles, but also showing how the particle description breaks down when too much energy is inserted into a finite area causal diamond.  It also leads to a resolution of the firewall paradox for evaporating black holes.  The formalism also allows one to construct a completely finite model of inflationary cosmology, with none of the usual Transplanckian puzzles, and an economical theory of dark matter, which may shed light on the early stages of galaxy formation.  

\section{QFT in Causal Diamonds}

In the 1960s, the inventors of Algebraic QFT discovered that the operator algebra constructed from bounded functions of fields smeared with test functions whose support lay in a causal diamond, was always a Type $III_1$ Murray-von Neumann factor in the algebra of all operators acting on the vacuum.  This is closely related to the fact\cite{sorkinetal} that the entanglement entropy of the diamond with its complement, in any cut off version of the theory, has a UV divergence proportional to the area of the holographic screen. It was proposed in\cite{sussugjacob} that this be viewed as a "renormalization of Newton's constant in the Bekenstein-Hawking law".  It was undoubtedly thinking about this generalization of the BH law from black hole horizons to general diamonds, that led Jacobson to his seminal paper\cite{ted95}.  The fact that the entropy of a diamond is finite in QG is an indication of the breakdown of QFT near the diamond boundary.

A more serious problem with the QFT picture of causal diamonds is that the logarithm of the number of states QFT assigns to a diamond scales like a power of the area greater than $1$.  If one computes the expectation value of the stress tensor in most of these states and uses it as a source in Einstein's equations in asymptotically flat space, then one creates a black hole larger than the diamond.  This is true despite the fact that one can keep the energy density far below the Planck scale.  The energy density cutoff has to scale like an inverse power of the size of the diamond in order to prevent the formation of supra-diamond black holes.   Once one imposes such a cutoff\cite{ckn}\cite{draperetal} the total entropy scales like {\it a power of the area less than 1}.  

Thus there seems to be a fundamental problem with thinking about the dominant contribution to the area law entropy of a diamond in a theory that contains black holes, as coming from QFT.  These problems are sharpened by the so-called {\it firewall paradox}\cite{ampsetal}.  Most of the states just inside the horizon, which contribute to the area law entanglement entropy, carry large local stress energy, {\it unless they are entangled more or less as they are in the vacuum, with states just outside the horizon}.  If a black hole evaporates to the point that it has lost half its entropy, Page's theorem\cite{page} tells us that the interior near horizon states must be entangled with the distant Hawking radiation.  One concludes that there must be large expectation values of the stress energy tensor, "a firewall" just inside the horizon.  Much paper has been consumed in trying to resolve this paradox without giving up the paradigm of a QFT description near the horizon.  We will instead suggest that, along with our previous remarks, we should really be thinking about a completely different formalism, with many fewer q-bits per causal diamond than QFT, and trying to understand how and when QFT emerges as an approximate description of this more holographic formulation of QG. 

The important paper of\cite{ckn} showed that not even the most stringent tests of QFT contradicted the ability to omit all states of QFT in the causal diamond defined by an experiment, whose back-reaction would have produced a black hole bigger than the diamond.  The authors of\cite{draperetal} showed that more sophisticated forms of the cutoff on field theory modes could put possible disagreement with experiment off into the distant future.  The essential point is that the experiment only probes space-time in a small region around the trajectory followed by the detector.  One needs to keep field modes up to the UV reach of the detector in that region, but outside of that region most short wavelength modes of the field cannot affect the measurement.  This would not be true if we tried to do simultaneous experiments in a large region of space, but then of course the mass of the detectors would collapse the region into a black hole.   

\subsection{The Hilbert Bundle Formulation and AQFT}

AQFT\footnote{I will discuss only the version of AQFT that describes things in terms of von Neumann algebras rather than $C*$ algebras.} assigns an operator algebra to every causal diamond in a Lorentzian space-time whose geometric structure is assumed in advance and is unconnected to the quantum dynamics.  The Hilbert Bundle formulation of QG makes a similar assignment, but the space-time is connected to the quantum theory by the relations we have outlined between quantum density matrices and diamond areas in empty diamond states. The background geometry is a hydrodynamic description of the quantum system.  We note that the requirement of finite entropy for a diamond does not by itself imply that the Hilbert space of the diamond is finite dimensional, since Type II factors can have density matrices with finite maximal entropy.  However, the fact that black holes can be formed by pure initial states over modest proper times, and that black holes in AdS space are thermal states of a quantum field theory in finite volume, suggest strongly that diamond Hilbert spaces are in fact finite dimensional.  In our discussion above, we implemented this requirement by coupling together a finite number of $1 + 1$ dimensional fermion fields, each of which had a fixed UV cutoff.  As we'll see, it's virtually impossible to imagine an experimental probe that can verify even the full details of a finite dimensional model, let alone prove that one really needs infinite dimensional local Hilbert spaces.  

AQFT also makes assumptions about the existence of a unique global state on a single Hilbert space on which all the individual local Type $III_1$ factors act.  This Hilbert space carries a unitary representation of the isometry group of the space-time.  In the Hilbert Bundle formulation, this assumption is replaced by the QPR.  There is no single Hilbert space on which all diamond algebras act.  However, in the case of asymptotically flat space, one can use the QPR to prove that the S-matrix is Poincare invariant.  This proof is in fact delicate, since it depends on an as yet incomplete understanding of the space of asymptotic states in asymptotically flat space time.  The conjecture\cite{tbags} is that it is a representation space for the Awada-Gibbons-Shaw\cite{ags} supersymmetric generalization of the BMS algebra.  The AGS operators are spinor valued half measures on the past and future null momentum cones, but the allowed states have to satisfy a collection of vanishing conditions for the spinor variables.  It has not been shown that these conditions define a separable Hilbert space. For asymptotically AdS space the path towards an understanding of symmetries is more clear, though no detailed calculations have been carried out.  We will discuss this in sections 3 and 4 below.

For an idealized model of dS space, the Hilbert bundle formulation has only the QPR. Although the global isometry group maps the fiber of the bundle over one geodesic into that over another, there are no actual "observables" for detectors on the two geodesics to compare, except those that can be captured by a finite time overlap.  In perturbative QG in dS space, each geodesic has gauge invariant observables associated with the other geodesics that enter into its horizon volume\cite{tbdsobs}.  However, all trace of these observables is erased by fast scrambling in a proper time of order $R_{dS} {\rm ln}\ (mR_S)$, where $m$ is the mass of the particle characterizing the perturbative observable, so there is no truly asymptotic data on which the global isometry group could act.  In the real world, the things "intelligent observers" can really measure depend on the existence of the complex gravitational bound states called local groups of galaxies, which define their own semi-classical reference frames where good detectors of quantum information can have lifetimes far exceeding the dS Hubble time.  These objects are contingent and idiosyncratic, and there can't be a universal symmetric mathematical theory of them.  At best one might come up with a statistical theory describing the probability distribution of observations from such a local group at late times.

It's also interesting to note the connection between the Hilbert Bundle formulation of QG and recent discussions of QG in terms of non-isometric encodings\cite{engeletal}.  One can view the map between the ensemble of Hilbert spaces of the bundle, and any single Hilbert space, asymptotic in proper time as time goes to infinity, as such a non-isometric encoding.  The QPR is the principle that ensures that all of these descriptions are exactly consistent with each other.  As we'll see, in the case of AdS space, our formalism meshes well with the tensor network description, which was the original inspiration for non-isometric encoding.  The main difference between the two approaches is that we do not pretend that QFT gives a good approximation within a causal diamond, and instead propose a more exact description of diamond local physics and the way in which field theoretic physics emerges from it for a certain set of low entropy constrained states.

Although we have argued that there is no reason to expect QFT to give a good description of most of the quantum information in a causal diamond, we must understand how the part of QFT that's been experimentally verified could emerge from our holographic description of the net of operator algebras describing space-time in QG.
The key to this was the Schwarzschild-deSitter black hole entropy formula, and a quantum mechanical implementation of it invented by Tomeu Fiol\cite{bfm}.  The SdS metric is
\begin{equation} ds^2 = - f(r) dt^2 + \frac{dr^2}{f(r)} + r^2 d\Omega^2 . \end{equation}
\begin{equation} f(r) = 1 - \frac{ r_s^{d-3}}{r^{d-3}} - \frac{r^2}{R_{dS}^2} . \end{equation}
\begin{equation} r_s^{d-3} = \frac{M}{(d-2)\pi A_{d-2}} ,  \end{equation} where $A_{d-2}$ is the area of the unit $d-2$ sphere.  This indicates that a mass inserted at the origin of a causal diamond in dS space {\it decreases} the entropy of the system.  If the Schwarzschild radius of the mass is small compared to the dS radius, then the decrease is
\begin{equation} \Delta S = \frac{M}{2\pi R_{dS}} , \end{equation} which mimics the effect of a Boltzmann factor at the Gibbons-Hawking temperature\cite{gh}.  This is a derivation of the GH temperature that is independent of QFT and of the fact that the Euclidean dS metric must have a periodic time if it is to be non-singular.

A similar calculation can be done when studying a small object of mass $m$ that is about to fall into a black hole in Minkowski space\footnote{Actually, this calculation works for any value of the c.c..  It is most dramatic when the black hole has negative specific heat, but it always gives the correct Hawking temperature.}.   The object will fall in and quickly, in a time of order $r_S {\rm ln\ } r_S$\footnote{For large black holes in AdS space, the infall time to the singularity goes like the AdS radius, rather than the Schwarzschild radius.}, form a larger black hole, with a huge increase of entropy, 
\begin{equation} \Delta S = \frac{m}{T_{Hawking}} . \end{equation} Conversely, by the principle of detailed balance, the probability for the black hole to spontaneously emit that object is given by a Boltzmann factor at the Hawking temperature.  

There are two lessons to be drawn from the common features of these examples.  The first is that a state in a causal diamond containing a large equilibrium system and an independent localized object is a constrained state of lower entropy than the equilibrium state of the whole diamond.  The second is that quantum field theory does not have any description of the additional degrees of freedom that must be "unfrozen" in order to reach the true equilibrium state of the diamond.  In all of these cases, even if we assume that QFT has a good description of the horizon, the state of "horizon + particle" is a low entropy excitation of the horizon itself.  From the point of view of the frame of reference falling with the particle, there are no other states of low energy that the particle could excite, to come into equilibrium with the horizon.  This is not true for an accelerated frame of reference that stays outside the horizon, but the QFT states that appear low energy to the accelerated observer, would contain high energy excitations ("a firewall") for the infalling object.  This argument can be sharpened\cite{ampsetal} using monogamy of entanglement for black holes that have evaporated away half of their quantum information.  Since we have already concluded that QFT has no reason to be a good description of the BHJFSB entropy of a diamond, these observations should just be viewed as a reinforcement of that point.

The essence of local physics is that localized objects cannot interact with each other when they are "far apart".  The structure of what is called "Einstein locality" in AQFT is built into our formalism by the use of time dependent Hamiltonians that generate the analog of "modular flows" in causal diamond algebras.  We  note that there is a theorem in QFT that exact causality of the type we've enforced {\it requires} Type III operator algebras if the Hamiltonian is time independent.  Our time evolution operators are the analogs of modular flows inside causal diamonds and are time dependent even on space-times that have global time-like Killing vectors, precisely in order to implement causality in a way that's compatible with finite dimensional local Hilbert spaces.

However, we also want to exhibit the kind of locality that is expressed by Feynman diagrams, where interactions inside a causal diamond come from a combination of Newton/Coulomb forces and "exchange of physical particles".  B. Fiol came up with the key idea that allows us to make models with these properties.  Consider a single trace matrix model with Hamiltonian
\begin{equation} N^{-p} {\rm Tr\ } P(M_A / N^k ) , \end{equation} where the $M_A$ are $N \times N$ matrices that are bilinear in fermionic oscillators and $P$ is a finite order polynomial
\begin{equation} M_i^j = \psi_i^{\dagger\ A} \psi_A^j  . \end{equation} If the index $A$ runs over $\sim N^{d-3}$ values, then the Hilbert space of this system has the entropy we expect of a causal diamond whose area scales like $N^{d-2}$.  This conclusion is unchanged if the $\psi_i^A$ are cut-off $1 + 1$ dimensional Dirac fields.  

Now imagine states that are constrained so that the matrices $M_i^j$ are block diagonal, with one block of size $\sim N$ and the others of size $\sim 1$.  Initially, the small blocks are independent systems, which have no interaction with each other or with the large block.  If $k = d - 3$ and $p = 1$, then the time scale for interaction is $o(N)$.  The number of fermionic oscillators that must be set to zero to make the off diagonal block between and small block and the large block vanish scales like a power of $N$.  Thus, even though the Hamiltonian is a fast scrambler, the equilibration time of the small block is of order $N {\rm ln}\ N$, so the small blocks remain independent systems during the entire time of propagation through a causal diamond of proper time $N$\footnote{We are simplifying the discussion a bit.  The actual Hamiltonian is time dependent, acting only on a subset of the fermions in each Planck time interval.  To discuss a scattering process, as we are implicitly doing, one should use a time symmetric nesting of proper time intervals. }.  

To get more precise agreement with space-time physics, we make the fermionic variables totally anti-symmetric tensors with $d-2$ indices.  The counting of variables is then the counting of angular momentum modes of a spinor on the $d-2$ sphere with angular momentum cutoff $N - 1/2$, which is asymptotically the same as the number of modes of the Dirac operator on any smooth Riemannian $d - 2$ fold with eigenvalue cutoff $ \sim N$.  The bilinears are then $p$ forms on the manifold, when multiplied by powers of the $d - 2$ bein, and the trace of polynomials of matrices is the approximate form of integrals of top forms over the manifold.  The manifest unitary invariance of the formula is a finite dimensional approximation to the group of volume preserving diffeomorphisms on the manifold.  

The variables $\psi_{ijk...}$ associated with a single small block interact with those having indices in the large $o(N)$ block in two ways.  They can form the bilinear matrix $M_i^J$ with the variables $\psi^{\dagger\ jk...\ J}$  and then this and its adjoint can be inserted between products of $M_i^j$ and $M_I^J$ to make non-zero traces.  Alternatively there are matrix elements of $M_i^j$ that can be made out of fermionic variables with one or more indices in the big block.  These are there even when the system is in a state where the bilinear matrix is block diagonal.  Since, as we'll see, the time scale for interaction of most of the $\psi_{ijk...}$ variables with themselves is much shorter than $N$, these terms can be viewed as a hint of the inevitable infrared problem of gravitational quantum physics.  The fact that the number of constrained q-bits necessary to suppress interactions scales like the $d - 3$ power of a variable associated with the length scale of a causal diamond should remind us of the formula for the Schwarzschild radius of a black hole in $d$ dimensional Minkowski space.  
So one should consider the $d - 3$ power of the size of the small blocks of the matrix $M$ to be the approximately conserved energy in a diamond.  It's conserved only in the infinite $N$, asymptotic limit.  

If the total number of constraints is such that most of the degrees of freedom are constrained, then the argument that the constraints are preserved for a time of order $N$ breaks down.  The system is then far from equilibrium and returns to equilibrium (no constraints) quite quickly.  If we look at the diamond of size $N$ as a subsystem of a much larger diamond, where we had the same total energy, but a much larger number of constraints, our diamond now looks like a single object of approximately fixed energy, in a high entropy equilibrium state, with an entropy proportional to its area.  That is, it has the properties of a black hole.    If we consider a time symmetric nesting of diamonds, appropriate for thinking about scattering experiments in asymptotically flat space-time, then starting from constraints with many fewer constrained q-bits than total variables, we will inevitably find situations where the number of constraints reaches the threshold we've associated with "black hole production".  So models of this type have asymptotic states with particle-like properties - the constraints can be followed on a time-like or null trajectory through a nested sequence of diamonds\footnote{More precisely our variables are like {\it jets} of particles.  They are areas of pixels on the holographic screen surrounded by an isolation zone of co-dimension one.}, which can create black hole like equilibrium states.   However, if the value of $N_{crit}$ at which the isolation between some number of small blocks in a diamond fails is not too large, the black hole formulas are not applicable.  The $N_{crit}$ diamond is constained in a larger one of some size $M$, which measured the total energy that went into it.  This must be conserved up to corrections of order $1/M$.  So the final state in the diamond $M$ is some number of particles carrying that amount of total energy.  To an effective field theorist with a resolution fuzzier than $M$, $M$ just looks like a multi-particle vertex.  In a large diamond, interactions between small blocks can thus occur on times scales much shorter than the proper time in the diamond by mergers of particle jets into such multiparticle vertices and (time-ordered ) Feynman diagrams in which a jet from one vertex propagates to another.  
\begin{figure}[h]
\begin{center}
\includegraphics[width=01\linewidth]{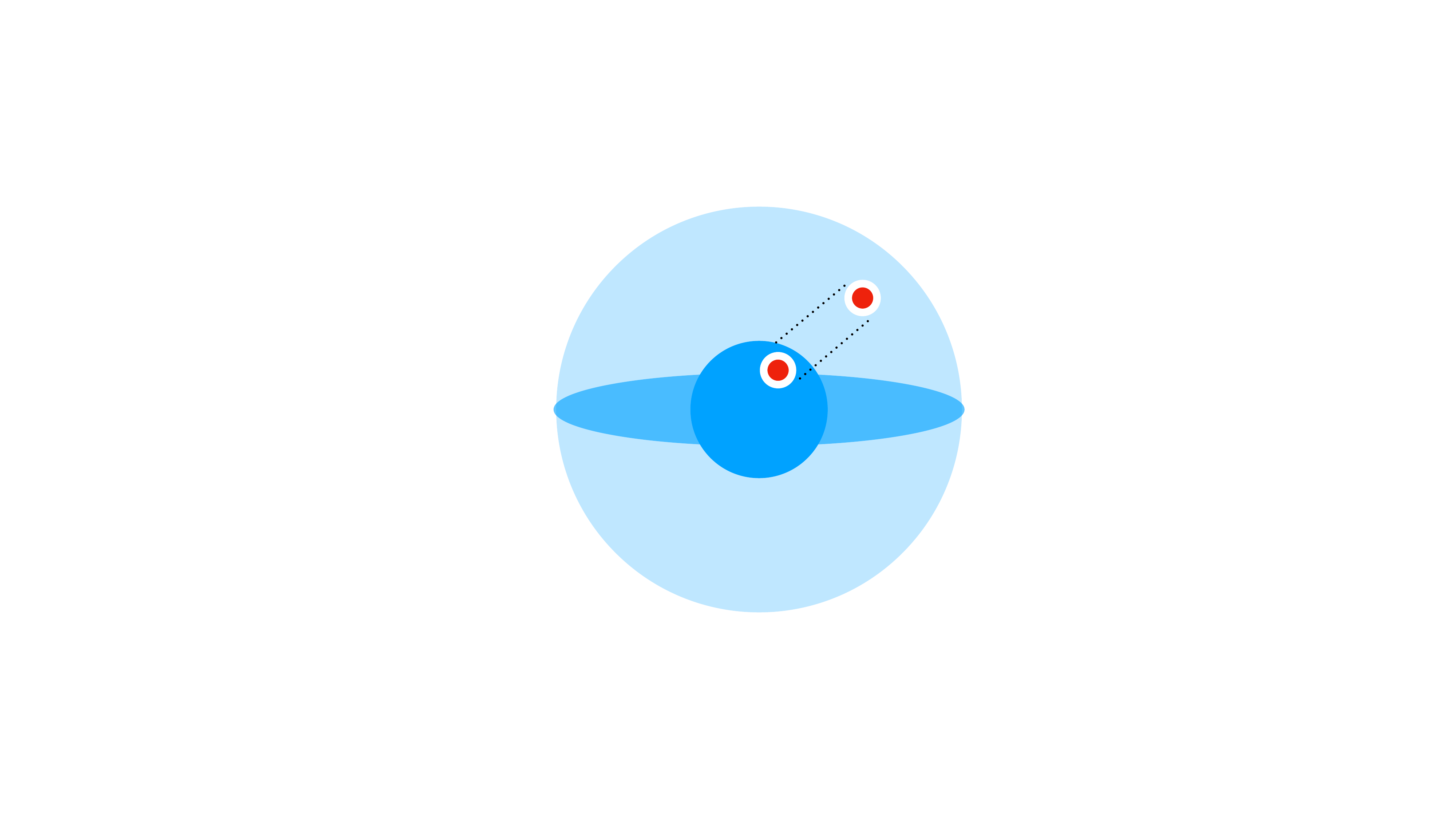}

\caption{The holoscreens of two consecutive nested causal diamonds, showing how following the constraints leads to a picture of the trajectory of a jet of particles through space-time. } 
\label{figure:layers}
\end{center}
\end{figure}

\begin{figure}[h]
\begin{center}
\includegraphics[width=01\linewidth]{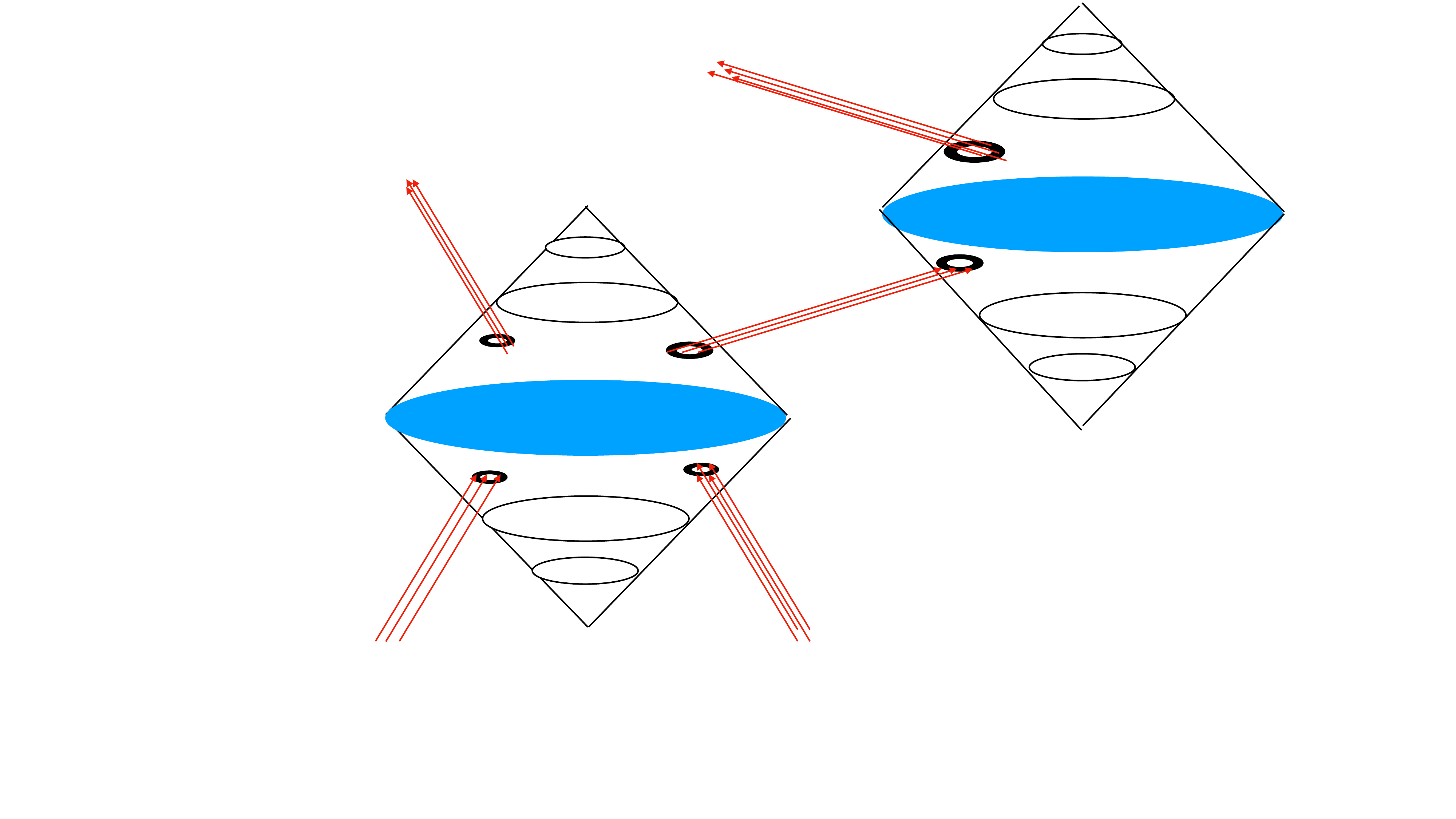}

\caption{Decomposition of amplitudes in HST models into time ordered Feynman-like diagrams, describing jets of particles propagating between causal diamonds in the background space-time. } 
\label{fig:feynman}
\end{center}
\end{figure}
Finally, one can estimate the leading long distance interaction between small blocks coming from virtually switching on and off the $\psi_{ijk...J}$ variables in second order perturbation theory\cite{tbwfnewton}.  It has the form of Newton's gravitational interaction.  It's thus plausible that models of the type described here can provide the clues to a general theory of quantum gravity.  The largest missing piece of the puzzle is a way to efficiently implement the QPR.  It's not clear whether the models described so far satisfy it.  

While these models have a long way to go before they become convincing, they definitely have a scattering theory whose asymptotic states can be described in terms of representations of the Awada-Gibbons-Shaw\cite{ags} supersymmetric generalization of the BMS algebra\cite{tbags}, and scattering processes that can involve both "local effective field theory vertices" and production of high entropy meta-stable equilibria with the scaling properties of entropy and energy of black holes.  Furthermore, the effective theory on "black hole horizons" and "empty causal diamonds" of the same area is identical.  The difference between the two subsystems of a larger causal diamond has to do with whether degrees of freedom not included in the small diamond are constrained.  This is consistent with the conclusion from classical general relativity than "there is no drama at the horizon".  In these models, the singularity encountered when a small object falls into a black hole is related to the equilibration of the object's degrees of freedom with those of the black hole.  It proceeds by excitation of variables that are not well modeled in the QFT description of the black hole horizon, but are instead captured by the Carlip-Solodukhin-Fiol matrix valued fermion fields on the stretched horizon.  

\section{Tensor Networks and Anti-deSitter Space}

Most of the work on holographic models of quantum gravity has been devoted to the AdS/CFT correspondence\cite{aharonyetal}.  This is a rather well established conjecture relating a class of quantum field theories to models of quantum gravity in space-times well approximated by versions of the Einstein-Hilbert Lagrangian.  All of these space-times have $d \geq 3$ dimensions that asymptote to AdS space with radius of curvature much larger than any microscopic length scale.  In all well established examples there are also at least two compact dimensions whose radius of curvature is of the same order as the AdS radius, and supersymmetry is restored in the UV spectrum of the theory\footnote{There is one notable class of exceptions\cite{evaetal}, in which a superconformal field theory in $1 + 1$ dimensions is perturbed by an exactly marginal Thirring interaction between R symmetry currents.  The limit of these models as the radius of $AdS_3$ goes to infinity is not understood.}.  In all cases, the c.c. in Planck units is a discrete parameter, which controls the asymptotic high energy spectrum of the model, rather than a calculated energy density of the lowest lying state.  This is a particularly striking confirmation of the conclusion from Jacobson's principle, that the c.c. should not be thought of as an energy density.  

Tensor networks evolved from both the Density Matrix Renormalization Group and the search to construct Quantum Error Correcting Codes.  They are a method for constructing ground state wave functions with the entanglement properties expected from short range lattice Hamiltonians.  Swingle\cite{swingle} recognized that if one put a tensor network on hyperbolic space, one obtained a systematic lattice cutoff of a conformal field theory.  This was put into practical use in the Tensor Network Renormalization Group\cite{tnrg} of Evenbly and Vidal.  The authors of\cite{harlowetal} showed how the ideas of error correcting codes explained puzzles about the reconstruction of bulk information from boundary correlators in AdS/CFT and Harlow\cite{harlow} demonstrated that conjectures\cite{RT}\cite{QES} became rigorous results about tensor networks.

Fischler and the present author\cite{tbwfads} interpreted the entangling maps of\cite{tnrg} as dynamics along geodesics in AdS space-time.  Consider closely packed spheres of some radius $R$ in $d - 1$ dimensional hyperbolic space.  The centers of the spheres sit on a lattice invariant under some infinite discrete subgroup of the isometry group of the space. With every sphere we will associate a Hilbert space whose dimension is $e^{A_{k} R^{k} / 4}$.  There is a unitary representation of the little group of the center of the sphere acting in this Hilbert space. Starting with some central sphere, which will be interpreted as the position of a time-like geodesic in $AdS_d$, the spheres organize into "shells".  The centers of the spheres in each shell form a $d - 2$ dimensional lattice.  The aim of the TNRG is to construct a sequence of lattice ground states, which converge on the ground state of some CFT as we go out to the boundary.
The technique used in\cite{tnrg} is to start with the Hamiltonian of some known critical lattice model with a small number of q-bits per site, and use the variational principle to construct entangling maps: unitary embeddings of one site on a particular shell to the Hilbert space of the few sites that it touches on the next larger shell.  This procedure clearly becomes scale invariant as one goes to large shell size.  The form of the entangling maps is determined by the variational principle, by minimizing the expectation value of the known critical model.  If instead, one starts off criticality, one finds that the procedure stops at a finite shell size, once the correlation length is exceeded.  Remarkably, at the critical point, the eigenvalues of the "entangling Hamiltonians" at quite small shell size, reproduce the spectrum of low lying dimensions in simple CFTs.  

In\cite{tbwfads} we interpreted the different shells of the tensor network as the holographic screens of causal diamonds corresponding to different intervals of proper time along our preferred AdS geodesic.  The relation between proper time and the tensor network shell is determined by the area/entropy relation, once we fix $R$ in terms of the AdS radius $R_{AdS}$.  This is a crucial point.  There is universal agreement, that for a model where $R_{AdS}$ is greater than all microscopic scales (we say such models have Einstein-Hilbert (EH) duals) $R$ must be of the order of $R_{AdS}$.  Furthermore, since all known models have $R_{AdS}$ size compact dimensions we must have $k \geq d$.   These results tell us that the understanding of local bulk physics provided by the tensor network construction does not probe locality below the AdS radius scale.  Conversely, since one can construct hyperbolic tensor networks for any CFT, the results of\cite{harlow} show us that properties like the Ryu-Takayanagi formula or the Quantum Extremal Surface formula, do not by themselve show that a model is well approximated by local bulk QFT.   Below the AdS radius, locality must be understood in terms of constraints on holographic q-bits, as outlined in the previous section.

Tensor networks provide a sequence of lattice approximations to a conformal field theory which make explicit the connection between the renormalization group transformations and time evolution along geodesics in AdS space.  As such, given the general lore of the renormalization group a properly defined TNRG for a CFT with an Einstein-Hilbert dual is guaranteed to exhibit all of the superconformal symmetries of the fixed point theory.  It is undoubtedly next to impossible to actually carry this program out in practice, given the complexity of the realistic models.  What is more plausible is to consider models that lie on a conformal manifold with an exactly soluble CFT, like ${\cal N} = 4$ Super Yang Mills, or the $AdS_3 \times S^3$ models that are continuously connected to permutation orbifolds.  One might build successful TNRG maps for the soluble models and this would make it more plausible that the same procedure worked in the more interesting regime where bulk gravity was a good approximation.  

\subsection{The Entropy of the Minkowski Vacuum and Poincare Invariance of the Scattering Operator}

In traditional massive QFT one assumes, and proves in perturbation theory and a variety of non-perturbative constructions, that there is a unique Poincare invariant vacuum state.
Conformal field theories generally also have a unique conformally invariant state.  Theories with spontaneously broken global symmetries have a moduli space of vacua parametrized by the coset space $G/H$, but the IR divergences associated with the S-matrix in those theories do not infect local correlation functions, and can be controlled by explictly breaking the symmetry.  The situation for massless spin one gauge bosons is more ambiguous.  The only case where we have a clear cut definition of a Hilbert space in asymptotically flat space is abelian gauge groups coupled to massive charged particles.  Nonetheless, all of these models can be studied rigorously by putting them in finite volume and trying to understand the infinite volume limit.

This is not the case for gravity.   We have known for many years that the Fock space formulation of gravity fails in four dimensions, but that it can be used to construct inclusive cross sections\cite{weinberg} that are finite to all orders of perturbation theory and sufficient for answering all experimental questions.  A Hilbert space formulation of the theory\cite{fadkul} has recently been called into question\cite{waldetal}.   In higher dimensions, there seem to be no perturbative problems with the scattering matrix and superstring models of various kinds appear to define sensible Fock space amplitudes to all orders in perturbation theory.  

However, arguments for both signs of the c.c. indicate that the zero c.c. "vacuum" is not a simple pure state.  Any theory of quantum gravity that accepts the area/entropy connection, has to conclude that a maximal causal diamond in empty dS space has a huge entropy.  In the limit of zero c.c., this seems to imply infinite entropy for empty space.  Our definition of localized excitations as constrained states of holographic q-bits is of course at odds with the Fock space picture and implies an infinite entropy for the Minkowski vacuum state.   However, since we have no agreed upon model of dS space in QG, these arguments are far from definitive.  In particular, the dS/CFT formalism postulates some kind of non-unitary CFT living on the space-like boundaries of global dS space as the proper arena for dS QG, though how the usual Hilbert space interpretation of quantum mechanics or the Minkowski S-matrix emerges from the dS/CFT correlators has never been explained.  There is no interpretation of the correlation functions of dS/CFT in terms of expectation values of operators in a positive metric Hilbert space.  Any such interpretation appears to be at odds with the claim that dS space has finite entropy. Recent results for infinite entropy 3d dS cosmologies\cite{cem} seem to indicate that even in this case, gauge invariant correlators need to be integrated over the spacelike boundary.  

The AdS/CFT correspondence is much more well defined, and here the flat space limit has been studied extensively, starting with\cite{polchsuss}.  The basic idea of that work is to take a CFT with an EH dual, and focus on a particular causal diamond called "the arena" whose size is large compared to all microscopic scales, but small compared to $R_{AdS}$.  Then one prepares a state in the past, a time-like distance of order $R_{AdS}$ from the past boundary of the arena, and acts on the vacuum with a finite number of local CFT operators.  One constructs those operators so that 
\begin{itemize}
\item The excitations created by them have high probability to enter into the arena as freely moving waves and do not encounter each other before penetrating the past boundary of the arena.
\item The final state of the interaction propagates out to the future boundary of the arena as well separated, freely moving waves.  In particular, if black holes are created inside the arena by the collision, their evaporation time is much shorter than the proper time in the arena.  
\end{itemize}
Then, it's argued that, in the limit $R_{AdS} \rightarrow\infty$ followed by $R_{arena} \rightarrow\infty$, the CFT correlators approach Minkowski S-matrix elements.  Perturbative calculations with Witten diagrams confirm these arguments, though it should be pointed out that, in the limit, one should be able to scatter things into the formerly compact directions of space, and no one has ever checked that this works properly.  This is particularly important for $AdS_3$ models, because there is NO S-matrix for gravity in three dimensional Minkowski space, so decompactification is crucial for these arguments to be valid.  

In\cite{tbwfads2} a problem with this construction of the S matrix was pointed out, a problem that is shared with perturbative calculations of the S-matrix directly in Minkowski space.   It order to prove that the S matrix exists as an operator in Fock space, it is insufficient to show that individual matrix elements between states with a finite number of particles exist.  One must show that the operator is unitary on Fock space.   In ordinary Minkowksi perturbation theory, there are finite amplitudes for production of states with arbitrarily large numbers of particles with arbitrarily low momenta.  In AdS space, there is a gap between the ground state and the lowest lying state, so for finite $R$ we cannot construct such states by the procedure of\cite{polchsuss} if the number of particles is large enough.

The clue to where soft graviton states come from is the observation that if $I_i (arena)$ are Polchinski-Susskind operators that create states inside the arena, as described above, and $O_k $ are operators that create bulk states localized outside the arena, then the connected correlators
\begin{equation} \langle I_i I_j \ldots O_m O_n \ldots \rangle, \end{equation} are non-zero, quite generically, for any finite $R_{AdS}$.  In the limit, these can contribute to matrix elements of the S-matrix with "arbitrary numbers of zero energy particles in initial and final states".  So it's clear that the full space of states on which the S matrix is a unitary operator in the limit need not be Fock space.

Another way to look at the same issue is to think about the reduced density matrix of the arena, in some state gotten by acting with operators $O_k$ on the CFT vacuum.  This is best done in the tensor network representation of states a fixed time along the geodesic between the past and future tips of the arena.  The state of the arena is the state of the central node of the network.  In the CFT vacuum that node is highly entangled with the much larger quantum system on the boundary, and all of its quantum information can be read off in many different ways from the boundary degrees of freedom\cite{harlowetal}.  Thus, its density matrix is close to maximally uncertain.  Acting with a finite number of $O_k$ designed to create bulk excitations localized far from the arena, will not change this conclusion.  So the empty arena density matrix, which converges to that of empty Minkowski space in the limit, has infinite entropy.   

Finally we note that similar conclusions can be drawn from the only non-perturbative formulation of quantum gravity directly in Minkowski space\cite{bfss}.  In this model, SUGRA scattering amplitudes are calculated as moduli space scattering amplitudes on the moduli space of a highly supersymmetric $N \times N$ matrix quantum mechanics, in the large $N$ limit with quantum energies restricted to $\sim N^{-1}$.  The moduli space consists of block diagonal matrices and there is conjectured to be a threshold bound state for blocks of any size $N$.  The bound state represents a super-graviton of longitudinal momentum $N/R$, where $R$ is the radius of the null longitudinal circle.   The light front energy of a state is $\frac{R \vec{P}^2}{N}$ and can be of order $1/N$ by having a large block, or a small one with $|\vec{P}| \sim \sqrt{N^{-1/2}} . $  The latter states represent "soft super gravitons", and one is clearly allowed to have amplitudes to emit an arbitrarily large number of them in the large $N$ limit.  In particular, there is a sector of the model in which there are no blocks of size $N$ at all.  This defines an infinite dimensional "soft graviton sector" of the Minkowski Hilbert space and it remains to be understood how the off diagonal block of the S operator between this sector and the Fock space of normal gravitons behaves in the large $N$ limit.  

To summarize, even in the highest dimension supersymmetric models we do not have a non-perturbative understanding of the Hilbert space of states of quantum gravity in asymptotically flat space or of the symmetry groups that act on it.  In particular, the role of the Bondi-Metzner-Sachs (BMS) algebra in dimensions higher than $4$ is completely obscure, and the Hilbert space in dimension $4$ is the most mysterious of all.  In\cite{tbags} I conjectured that the proper formulation was in terms of a representation space of a generalized Awada-Gibbons-Shaw\cite{ags} algebra, the supersymmetric generalization of the BMS algebra.  The generators are $Q_a^i (P), \tilde(Q)_a^i (P)$, which live on the null cone in $d$ dimensional Minkowski space.  They transform as spinors under Lorentz transformations and satisfy left moving and right moving constraints w.r.t. the null momentum $P$, respectively.  The index $i$ has to do with compact dimensions of space, and the commutation relations between left moving and right moving generators encode information about the masses of stable massive particles.  There are separate generators for the positive and negative energy null cone, which are identified with the remote past and future of the Minkowski Penrose diagram.  These generators are mapped into each other by the S operator.  The magnitude of the null cone momentum is Fourier dual to the null coordinate on null infinity, and can also be thought of as the spectrum of the BMS generators.  

The generators are half measures on the null cone and the space of states is defined by an intricate series of constraints.  For finite $P(1, \pm \vec{ \Omega})$ the generators vanish unless the angles are constrained to lie inside of a finite number of non-overlapping regions.  Furthermore, at $P = 0$ the generators vanish in "annuli" surrounding those finite regions.  These are supposed to mimic the criteria defining Sterman-Weinberg jets in QCD.  From the point of view of the matrix models in this paper, these constraints arise from the vanishing of off diagonal matrix elements, as outlined in\cite{hilbertbundles}.  While it is fairly clear that one can take a limit of the finite systems discussed in this paper that gives this space of states, it is less clear that the resulting limit defines a separable Hilbert space.  In order to prove that the resulting S operator is Poincare invariant, one would have to show that the dynamics outlined in this paper satisfies the QPR.  Perturbative string theory suggests that this only works for special choices of compactification manifold {\it etc.}, so there is probably a lot of work to do to find completely consistent models.  

To summarize, in the Hilbert Bundle formulation, isometries of the background hydrodynamic manifold generally map one fiber of the Hilbert bundle to another and do not act on any single Hilbert space.  They {\it might} act on the asymptotic Hilbert space of a given fiber in the limit that its area goes to infinity.  This has been established, at least in principle, for the case of negative cosmological constant by the technique of holographic renormalization of tensor networks.  For vanishing cosmological constant, the non-perturbative answer to this question awaits the clarification of the nature of the asymptotic Hilbert space.  For positive cosmological constant the symmetry group does not act.  Ironically, as we will see below, this does not interfere with the approximate de Sitter invariance of low point correlation functions in inflationary cosmology, because inflationary cosmology is not de Sitter space.  

These remarks point up the role of the cosmological constant in a theory of quantum gravity.  It is not, as Jacobson taught us, a contribution to the stress energy tensor.  Rather it regulates the relation between the two parameters, {\it area} and {\it proper time}, which control the asymptotic size of a causal diamond.  As such, when the concept of energy makes sense, it also controls the behavior of the highest energy spectrum in the model.  Different models of QG have different values of this parameter, and it is not affected by renormalization in any low energy effective field theory approximation to the model.  

\section{Low Dimensional Models}

The BHJFSB area law makes no sense in $ 1 + 1$ dimensions, and, not coincidentally, the EH action becomes purely topological.  Dimensional reduction of higher dimensional models makes it clear that what one needs is a scalar field that plays the role of entropy in a causal diamond, which multiplies the EH term
\begin{equation} I = \int\ S \sqrt{-g} R . \end{equation}  One can then invent a parallell to Jacobson's 1995 argument\cite{bdz} and conclude that the general long distance action for gravity is
\begin{equation} I = \int\ \sqrt{-g}[S R  - \frac{1}{2} ( \nabla S)^2 - V(S)] . \end{equation} We've performed Weyl transformations on the metric to standardize the coefficient of the kinetic term of $S$.
These are called dilaton gravity models.  

The CS ansatz tells us that we should expect the quantum systems corresponding to these hydrodynamic equations to be cut off two dimensional CFTs, with central charge matched to the classical value of the S field in a diamond.  Note that dilaton gravity models have no field theoretic degrees of freedom.  The $S$ field always has level surfaces on which it doesn't vary.  We can always choose the coordinate along which it does vary to be locally orthogonal to those.  The equations of motion are
\begin{equation}   \nabla^m \nabla^n S - g^{mn} \nabla^2 S + g^{mn} S - \frac{1}{2} \nabla^m S \nabla^n S - \frac{1}{4} g^{mn} (\nabla S)^2 - \frac{1}{2} g^{mn} V(S) = 0 . \end{equation}
\begin{equation} - \sqrt{-g}^{-1}\nabla (\sqrt{-g} \nabla S) + V^{\prime} (S) - R = 0 . \end{equation} 
Call $t$ the coordinate direction along which $\partial_t S = 0$.  Then we have
\begin{equation} - g^{tt} \nabla^2 S + g^{tt} S - \frac{1}{4} g^{tt} (\partial_r S)^2 - \frac{1}{2} g^{tt} V(S) = 0 . \end{equation}  This is
\begin{equation} V(S) + 2 S + g^{rr}[\partial_r^2 - \partial_r ({\rm ln}\ g_{rr}) \partial_r ]S   + \frac{1}{2} (\partial_r S)^2 = 0 . \end{equation}
The equations with mixed indices vanish identically in our gauge choice, while the $rr$ component is 
\begin{equation} [\partial_r^2 - \partial_r ({\rm ln}\ g_{rr}) \partial_r ]S (g^{rr}) +  S - \frac{3}{4} 
(g^{rr}) \frac{1}{2} (\partial_r S)^2 - \frac{1}{2}  V(S) = 0 . \end{equation}
These two equations determine $S$ and $g_{rr}$ as functions of $r$ and imply that $g_{rr}$ is independent of $t$.  The equation of motion for $S$ then determines $g_{tt}$, which is also independent of $t$.  The causal structure of the solutions is determined by the possible sign changes in $g_{nn}$.  $g_{tt}$ is chosen to be time-like at infinity, and $g_{rr}$ space-like.  

In two dimensional gravity then, the general message of Jacobson's 1995 paper is particularly clear: the gravitational variables cannot account for the entropy implied by the values of the $S$ field on the boundaries of causal diamonds.  We must construct quantum systems that match those predictions.  However, the CKN arguments also fail in two dimensions, and indeed the CS refinement of Jacobson's principle leads us to expect that cutoff CFT is indeed the right arena in which to search for models of two dimensional quantum gravity.  

In fact, a somewhat earlier and more rigorous line of argument leads to precisely the same conclusion.  This is the famous construction of models of two dimensional gravity from string theory and double scaled matrix models\cite{liouvillematrix}.  Let us recall how this works in the simplest case of Type 0B string theory in the linear dilaton background.  This is the RNS string with SUSY violating boundary conditions in two dimensional Minkowski space time with a string dilaton profile
\begin{equation} g_s^2 = e^{- r/L_s}  . \end{equation}  
This space-time has a null singularity where $g_s^2$ blows up and the dilaton $S = g_s^{-2}$ goes to zero.  The detailed description of Type 0B string theory and its relation to the matrix model we are about to discuss can be found in\cite{newhat}\cite{maldseib}.

One starts from a generic model of $N \times N$ hermitian matrices with a potential term
${\rm Tr}\ V(M)$ that has a maximum at $M = 0$.  The unitary conjugation symmetry
$M \rightarrow U^{\dagger} M U$ is gauged, so that only unitary singlets, the matrix eigenvalues, are physical variables.  The measure over the eigenvalues famously is the van der Monde determinant
\begin{equation} \prod_{i \neq j} (\lambda_i - \lambda_j)^2 . \end{equation} The residual permutation gauge symmetry tells us to treat the $\lambda_i$ as identical bosonic particles, but if we absorb half of the van der Monde determinant into the boson wave function, this turns them into fermions.  

The double scaled limit of the model is obtained by taking $N \rightarrow \infty$ and letting the fermi level approach the top of the potential as $1/N$.   The result is a non-relativistic quantum field theory of fermions in an upside down oscillator potential, which has been analyzed rigorously by Moore\cite{moore}.  A non-perturbatively stable definition has two parameters, the energy scale defined by the curvature of the potential, and the ratio of the distance between the re-scaled fermi level and the top of the potential.  The observables are scattering amplitudes, which we will see in a moment can be viewed as amplitudes for scattering of two species of relativistic massless Dirac fermions on a half space, with a Minkowski metric, a linear dilaton profile, and a "tachyon wall background". When the fermi surface is far below the top of the potential, there is a perturbation theory which, after appropriate identifications, matches the Type 0B string perturbation series.  Linear combinations of the $U(1)$ currents associated with the two fermions are derivatives of the NS and RR "tachyon" fields of the low energy effective action of the string theory.  The tachyons are in fact massless because of the linear dilaton background. To match the two computations of scattering amplitudes one must make the proper momentum dependent normalization of string theory vertex operators\cite{Yin}.  This is related to the Castillejo-Dalitz-Dyson ambiguity, which is particularly severe for massless scattering in $1 + 1$ dimensions.  

It is easy to see the transformation to relativistic fermions, by writing the Hamiltonian as\cite{AKK} and the appendix to\cite{maldseib}, 
\begin{equation} H =  \int d\lambda\ \frac{1}{2} \psi^{\dagger} (\lambda) (p_{\lambda}^2 - \lambda_2) \psi (\lambda) = \int d\lambda_{\pm} \psi^{\dagger} (\pm \lambda_{\pm})[(-i \lambda_{\pm}\partial_{\lambda_{\pm}}  \pm i ]\psi ( \lambda_{\pm}) . \end{equation}
\begin{equation} \lambda_{\pm} = p_{\lambda} \pm  \lambda , \ \ \ x_{\pm} \equiv {\rm ln}\ (\lambda_{\pm} . \end{equation}
\begin{equation} H = \pm i \int dx_{\pm}\ \psi^{\dagger} (x_{\pm}) \partial_{x_{\pm}} \psi (x_{\pm}) . \end{equation} 
The single particle wave functions are singular at $\lambda = 0$, and careful considerations\cite{AKK}\cite{maldseib} show that there are two species of Weyl fermions, with S-matrix
\begin{equation} \Psi_{out} (r,t) = \int_{-\infty}^{\infty}ds\ e^{\frac{1}{2}(r + s)} [e^{i(r + s)} + e^{-i(r + s)} \sigma_1]\Psi_{in} (s,t). \end{equation} The barrier appears as a chemical potential for the relativistic fermions.  
When the fermi level is far below the top of the barrier, linear combinations of the two species correspond approximately to non-relativistic fermions on either side of the potential, to all orders in perturbation theory.   One can then compare this perturbation theory to string theory and Yin and collaborators have shown\cite{Yin} that with an appropriate definition of the vertex operators, the two scattering amplitudes coincide up to second order.  More generally, the correct linear combinations are defined by symmetry and are the even and odd combinations under reflection in $\lambda$.  

Despite the correspondence to string theory and a low energy effective description including the gravitational field, this model is completely integrable.  Although the low energy effective action has black hole solutions, there are no states of the quantum theory that behave like black holes.  This is particularly clear because of complete integrability\cite{maldaetal} but there is a more fundamental reason.  All of the states of the model are concentrated in the region where the excitations behave like free relativistic fermions.  The strongly coupled region corresponds, in the non-relativistic picture, to the region near the maximum of the potential, where the density of states goes to zero.  Despite these deficiencies, the correspondence between the dilaton field $S = g_s^{-2}$ and entropy, does work in these models\cite{dasetal}.

We can fix all of these problems simultaneously by a pair of modifications.  We first increase the number of fermion fields to a large number $M$, and then we add interactions that are concentrated near the top of the barrier.  First we add 
\begin{equation} H_+ = - \int d\lambda f(\lambda) [\psi_i^{\dagger} (\lambda)\psi_i^{\dagger} (\lambda)]^2, \end{equation} where $f(\lambda)$ is exponentially localized near the origin.  
If we send in a pulse with a large expectation value of $[\psi_i^{\dagger} (\lambda)\psi_i^{\dagger} (\lambda)]$, this will generate a self consistent attractive potential that traps a large number of fermions near the origin.  In addition we add an SYK\cite{SYK} like coupling, chosen from a random distribution
\begin{equation} \int dx dy dz dw g_{ijkl} \psi_i^{\dagger} (x) \psi_j (y) \psi_k^{\dagger} (z) \psi_l (w) \langle x | e^{-\beta (p_{\lambda}^2 + \lambda^2)} | y \rangle  \langle z | e^{-\beta (p_{\lambda}^2 + \lambda^2)} | w \rangle . \end{equation}

If we analyze these models far from the maximum of the potential, and use the AKK map, we find a model of $2M$ massless relativistic fermions propagating in a Minkowski background, with both incoming and outgoing waves.  If we take the Fermi surface far below the maximum of the potential we can compute a perturbative S matrix, which for scattering of $\ll M$ particles is just the tensor product of $M$ copies of the $0B$ S-matrix. 

Taking the Fermi surface to lie right at the maximum of the potential, we are in the strong coupling regime, but the asymptotic region is still described by free fermions.  It's clear that we should identify the $S$ field with the total entropy of the fermions, which implies, via the matching formulae of\cite{dasetal}\cite{lindil1} that we are in the linear dilaton vacuum of the CGHS model\cite{CGHS}.  This model was originally introduced to describe the generic throat region of 4 dimensional charged linear dilaton black holes in Type IIB string theory, and the massless fermions can be thought of as originating with massless Ramond-Ramond fields in a Calabi-Yau compactification.  One might imagine that the random couplings we have introduced reflect the many different types of charged linear dilaton black holes that can be obtained in such string compactifications.
The Lagrangian of the CGHS model is
\begin{equation} {\cal L} = (2\pi)^{-1} \int d^2 x\ \sqrt{-x} [e^{-2\phi} (R + 4(\nabla\phi)^2 + 4\lambda^2) - \frac{1}{2}(\nabla f)^2]    . \end{equation}

There is thus evidence from the Type 0B string that we can view our modified models as having a gravitational dual, which behaves like linear dilaton gravity asymptotically.  Unlike the Type 0B string, our models are not completely integrable, and they have only a single conservation law, associated with the total $U(1)$ charge $\int d\lambda \psi_i^{\dagger} (\lambda) \psi (\lambda) $.  We conjecture that this is identified with the charge of the linear dilaton black holes of the CGHS model.  The apparent global $SU(N)$ symmetry of the perturbation series is broken by the random couplings, as are all of the symmetries associated with powers of the single particle Hamiltonian $p_{\lambda}^2 - \lambda^2$.   

For large $M$, the system has long lived meta-stable excitations, which can be created by sending in scattering states with large localized values of $ \langle \sum_i \psi_i^{\dagger} (\lambda) \psi_i (\lambda) \rangle $.  Using the time dependent Hartree approximation near the origin, the attractive four fermi interaction creates a confining potential, which is the same order as the kinetic energy when the expectation value is large.  The random couplings will then act to thermalize the state, on a time scale of order $1$, the scale set by the kinetic energy.  In making these statements, we use the usual rules of large $M$ physics, which scale the couplings to make the interaction terms comparable to the kinetic terms as $M \rightarrow \infty$.  
If we now throw another fermion into the meta-stable excitation, it thermalizes in a time of order $1$.   Similarly, the time for a fermion to tunnel out of the confining potential so that it can escape to infinity is of order $1$.

Let us compare these statements to the predictions of linear dilaton gravity for linear dilaton black holes\cite{CGHS}.  These are high entropy meta-stable states, which can be created by sending in pulses containing a high density of fermions from the asymptotic region.  The region inside the horizon is microscopically small, in the sense that an observer hits the singularity in a time that's independent of the black hole entropy.   Finally, the Hawking temperature is also independent of the black hole entropy.  These properties exactly mirror those of the meta-stable states of the model that we described above, if we identify the phrase "hitting the singularity" with "equilibrating the infalling system with the degrees of freedom of the black hole".   Note that this matches the correspondence between infall time and inverse frequencies of quasi-normal modes that one observes for higher dimensional black holes.  

Since our model is a unitary quantum system, the evaporation of our meta-stable states will follow a Page curve.  The "island formula"\cite{islands} does not really make much sense for this system, since the region behind the horizon is always microscopically small and effective field theory is never applicable there.  The semi-classical analysis of\cite{thunderpop} does however give a reasonable approximate description of the evolution\cite{lindil2}.  Finally, note that we can use the AKK transform to rewrite our entire Hamiltonian in terms of relativistic Dirac fields.  In this language, the interaction terms look highly non-local and ugly and the physics is far from transparent.  

Another famous dilaton gravity model is Jackiw-Teitelboim gravity with negative cosmological constant.  This is supposed to model the near horizon region of extremal $4$ dimensional Reissner-Nordstrom black holes of large charge.  The four dimensional geometry has the form  $AdS_2 \times S^2$ but the condition of extremality does not allow for angular momentum excitation so all the states are spherically symmetric.  The standard approach to this model views it as exploring the region just {\it outside} the horizon and claims that the quantum mechanical dual is a certain limit of the SYK\cite{SYK} ensemble of models.  The problem with this approach is that it does not really exhibit a large radius $AdS_2$ space-time.  JT gravity has a boundary Schwarzian mode, related to the breaking of $SL(2,C)$ symmetry and the limiting SYK model certainly has precisely the same mode.  However, there's no indication of massless propagating modes in $AdS_2$ and the excitations that exist appear to have "string scale" masses.

There is an alternate approach\cite{bdz}, based on the AKK transform, which is less familiar and does exhibit a large radius $AdS_2$ in keeping with our expectations from the classical geometry of RN black holes of large charge.  We can understand the physics we expect to find by studing the bosonized fermion current of a massless Dirac field propagating in the $AdS_2$ black hole geometry of JT gravity.  The geometry is a black hole because the classical solution for the entropy field in global $AdS_2$ coordinates is 
$$  S = \Phi_h \sqrt{1 + x^2} \cos \tau   .$$   We cannot allow the entropy to fall below zero.  This restricts the global $AdS_2$ geometry to the regime covered by the coordinates
\begin{equation}  ds^2 = - (r^2 - r_s^2) dt^2 + dr^2 /(r^2 - r_s^2) .\end{equation} \begin{equation} \Phi = \phi_b r , \end{equation} with $r \geq 0$, which has a Killing horizon for the generator of asymptotic time translations.  These are the coordinates in which the dilaton/entropy field is time independent, outside the horizon.  Fermionic current flowing out of the horizon reaches the boundary at infinity at finite time.  If we are modeling a stable extremal RN black hole, this current should be identified with the electric current of the hole and charge conservation implys that it should reflect off the boundary and fall back to the horizon, in a time of order $R_{AdS}$.  For large charge, the AdS radius is much larger than the Planck scale or any other macroscopic scale.  

Standard quantization of massless fermion field theory coupled to gravity, using Schwinger's prescription of matching Lagrange to Heisenberg equations of motion to determine the commutation relations, leads to the following conclusions\cite{bdz} for states involving coherent flows of current in the AdS region:
\begin{itemize}
\item Current, as well as energy and momentum flows out of the horizon by Hawking radiation, at constant temperature.  It bounces off the boundary at infinity and returns to the the vicinity of the horizon in a finite time, but never quite reaches it because of the familiar redshift.
\item The value of $S$ near the horizon always increases due to the flow of energy into the near horizon region and decreases do to outflow of energy. Interpreting the field $S(r,t)$ as the entropy of a causal diamond with one tip on the boundary at infinity and the other at a position $r$, this leads to a formula for the entropy in diamonds in equilibrium states
\begin{equation} S_{eq} (r) = S_0 - \mu^2 (r - r_s) , \end{equation} where $r_s$ is the position of the horizon.  The constant in this formula gets an infinite additive renormalization in field theory.
\item The last remark together with the finite entropy of RN black holes, suggests that the proper theory of QG is obtained by simply cutting off the QFT.  We'll see things are a bit more complicated.  

\item We consider {\it only} positive $\mu^2$ whereas most discussions of JT gravity take $\mu^2$ negative and claim to be studying near extremal perturbations of the extremal RN black hole as in conventional AdS/CFT correspondences.  We find that point of view suspect, in view of the results of\cite{mms} and the fact that the proposed SYK dual does not exhibit a large radius $AdS_2$.  Our model describes only transient excitations of the extremal black hole, which are invisible to an external observer, but involve localized degrees of freedom Hawking radiated from the $AdS_2$ horizon and bouncing off a barrier at the boundary of the $AdS_2$ region.  There appears to be a one parameter family of meta-stable equilibria, parametrized by $\mu^2$.  
\end{itemize}

Like all two dimensional Lorentzian geometries, $AdS_2$ is locally conformal to Minkowski space.  Thus, we can always map the massless Dirac equation on $AdS_2$ to that on flat space, and via the AKK map, to the upside down oscillator.  Since we want to match the time of the coordinate system where $S$ is static to the time of our time independent Hamiltonian system, the boundary conditions on the non-relativistic fermions are very different than those we used for the linear dilaton system.   Classical motion reaches the top of the potential in finite time, so we identify this with the asymptotic infinity of $AdS_2$ and impose a reflecting boundary condition there.  The oscillator is only one sided.  The classical travel time to infinity is logarithmically infinite, which coincides with the behavior of infall to the horizon in static coordinates, so we identify the horizon with infinity in the oscillator coordinates.  

The combination of the AKK map, and the conformal map between static $AdS_2$ coordinates and Minkowski space now allows us to map any single fermion motion or coherent current flow from the space-time picture to the inverted oscillator picture.  What we need to do to finish the correspondence is to find a non-relativistic fermion state that matches the properties of the static $AdS_2$ black hole of $JT$ gravity, in the sense that it assigns the same entropy to each causal diamond.  This was accomplished, approximately, in\cite{bdz}.  The two Hermitian operators $p_{\lambda} \pm \lambda$ satisfy exact quantum equations of motion with rising and falling exponential solutions.  We can build a density matrix from functions of these, and use the approximation of replacing Moyal brackets with Poisson brackets to evaluate expectation values and entropies of finite regions.  It's then straightforward to cook up a density matrix that reproduces the classical space-time entropy formula above, for each causal diamond, using the maps between the oscillator coordinate and the space-time coordinates.  Note that these maps are all completely local functions of the spatial coordinate only.  

In order to render everything finite and destroy the integrability of the model, we have to cut off the spatial extent of the $\lambda$ coordinate and add multi-fermion interactions. This can be done by recalling the origin of the model as the double scaled limit of a matrix model.  The large $\lambda$ region is replaced by the minimum of the original stable matrix model potential, and we can localize all of the multi-fermion interactions there.   From the point of view of the space-time picture, this is the region "behind the horizon" where the space-time interpretation of quantum gravity breaks down.  Again, as in the linear dilaton model, this region does not have a macroscopic extent in the classical geometry.  

In summary, we've constructed two families of quantum models that appear to be dual to classical dilaton gravity actions in $1 + 1$ space-time dimensions, in the sense that the hydrodynamics of high entropy states of the models matches the classical gravitational dynamics of dilaton gravity.  In the linear dilaton model this includes the production and decay of black holes, while in the JT gravity model one describes only the flow of entropy, charge and energy in the macroscopic $AdS_2$ throat.  In both cases, the quantum models go beyond hydrodynamics and match neatly on to the semi-classical dynamics of massless fermion fields coupled to dilaton gravity via minimal coupling to the metric.  Both classes of models admit a large number of random couplings.  We've speculated that the origin of this plethora of different models owes its origin to the many types of black hole throats that can be constructed in consistent 4 dimensional superstring models.  

A problem we have not attempted to solve is how to extend the model of extremal RN black holes to near extremal holes.  This is purportedly accomplished by the standard mapping of negative $\mu^2$ solutions of $JT$ gravity to the SYK model.  We have already noted our skepticism about this claim.  Certainly no one claims to have the kind of mapping between four dimensional scattering amplitudes from the black hole and computations in the SYK model that one usually obtains from the AdS/CFT correspondence for higher dimensional AdS spaces.   

We emphazise that both our general matrix model technology and the Hilbert bundle philosophy fail in these models, but the general ideas of classical gravity as a hydrodynamic description of quantum gravity and the Carlip-Solodukhin ansatz survive.  The matrix model technology was invented to deal with the transverse geometry of diamonds, which doesn't exist in $1 + 1$ dimensions.  The Hilbert bundle idea proves less useful for a different reason.  Both of the systems we've studied have preferred geodesics, where all the interesting physics occurs.  It is singled out by the classical behavior of the entropy field.  In either of these systems, studying causal diamonds other than those centered around infinity or the horizon/singularity, is an awkward complication.  

  \section{Cosmology}
  
  Consider a flat four dimensional homogeneous isotropic universe in which we assume we're in the empty diamond state for all times, but that the growth of the area of the horizon asymptotes to a finite value.  According to the CS principle, as generalized by\cite{BZ}, the entropy grows like
\begin{equation} S = (t/L_P)^2 \lambda L  , \end{equation} where $\lambda$ is the ultraviolet energy/momentum cutoff in the two dimensional CFT and $L$ is the interval on which it lives.  Conventionally, we measure $L_0$ eigenvalues as pure numbers and set $L$ equal to a multiple of $\pi$, but we want to introduce energy as measured along the geodesic in the diamond.  The important point is that $\lambda L$ is independent of $t$.  From the point of view of a detector on the geodesic, the states on the cosmological horizon have energy that scales like $1/t$.  Thus the expectation value of the energy in our state scales like $t$.   The volume of the horizon scales like $t^3$ and so the energy and entropy densities scale like 
\begin{equation} \sigma \sim \sqrt{\rho} \sim t^{-1} . \end{equation}
Thus the CS ansatz predicts a $p = \rho$ equation of state as long as the universe keeps expanding and is flat.  If the universe has negative curvature, the CS ansatz breaks down for finite area diamonds whose size is larger than the radius of curvature\cite{BZ}.  For positive curvature the universe recollapses when the size of the diamond gets to the radius of curvature.  Fischler and Susskind\cite{fs} and Bousso\cite{bousso} showed that the covariant entropy bound cannot be saturated in these cases.  

We can insist that the expansion of the horizon area asymptotes to a finite value by adding a cosmological constant to the $p = \rho$ model.  The scale factor is
\begin{equation} a(t) = \sinh^{1/3} (3t/R_I) , \end{equation} where $R_I$ is the Hubble radius associated to the c.c. .  This universe saturates the covariant entropy bound at all times and leads to a model of the dS horizon consistent with the CS ansatz\cite{tbpdreplica}.  This universe has no localized excitations in it at any time, and as a consequence it is very easy to satisfy the QPR.   

The point is that because all we know about the subsystem in the intersection is the size of its Hilbert space and that it is a subsystem of a (typically much larger) system that is a fast scrambler\cite{lshpss} we can expect the density matrix to obey the generic CS ansatz, the only general constraint that Einstein's equations put on models of quantum gravity. In principle this could allow the modular Hamiltonians of this diamond to be the $L_0$ generator of the CFT at a different points in the conformal manifold emanating from the same set of free fermions, but that would be incompatible with the modular flows in the individual diamonds, which are identical to each other.  Note by the way that although our background geometry is homogeneous, isotropic and flat, there is no fine tuning of initial conditions in our quantum system.  

To see that this cosmology admits no localized excitations we can turn to the final state.  A localized excitation in dS space disappears from any maximal causal diamond classically, unless it is traveling along the diamond's geodesic.  On the geodesic, its distant gravitational field looks like that of a black hole of the same mass
\begin{equation} ds^2 = - f(r) dt^2 + dr^2 / f(r) + r^2 d\Omega^2 . \end{equation}
\begin{equation} f(r) = 1 - 2ML_P^2 / r  - r^2 /R^2) . \end{equation}
When the Schwarzschild radius is $\ll R$, this gives an entropy deficit $2\pi M R/\hbar$ which is what one expects from a Boltzmann factor at the Gibbons Hawking temperature.  This tells us that localized states are states with of order $R/L_P$ constrained q-bits, compared to a generic state in the CS ensemble.  The behavior of states on generic trajectories and of quasi-normal modes of the horizon suggest that the equilibration time is of order $R {\rm ln} (M L_P) . $.  

\section{The Holographic Inflationary Cosmology}

The Holographic Inflationary Cosmology (HST) takes a collection of the previous section's maximal entropy models and incorporates them into a larger model in which the horizon slowly expands but the individual maximal entropy models remain as approximately non-interacting subsystems of the larger model for some period of time.  To see how this can be done, we use Fiol's matrix model realization of the de Sitter black hole entropy formula for four dimensional dS space, updated to include the CS ansatz.  The spinor bundle on the two sphere has a basis that can be viewed as the $N \rightarrow \infty$ limit of the set of $N \times N + 1$ matrices $S^i_A$, the spinor spherical harmonics up to some maximal spin.  We introduce massless two dimensional Dirac fields $\psi_i^A (\tau, \sigma) $ on the stretched cosmological horizon at time $t$, which are the expansion coefficients of a spinor on the two sphere into harmonics with $t = N$.  These represent the fluctuations of the geometry of the holographic screen around the spherical geometry of the hydrodynamic FRW background.  We do this along each geodesic of the cosmology independently.  We will discuss things from the point of view of a single geodesic and then study whether our proposal is consistent with the QPR.

To create an initial state on the past boundary of some causal diamond in which there is no interaction between subsets of degrees of freedom, we sandwich the empty diamond density matrix with a projection operator imposing constraints
\begin{equation} \psi_i^{A\ \dagger} | S \rangle = 0 , \end{equation}  for some collection of matrix elements where the $i$ index runs over $R_I/L_P$ values and the $A$ index runs over $R/L_P$ values.  Here we are assuming that the universe ends up in a de Sitter state with $ R \gg R_I$ and imposing boundary conditions on the past boundary of the maximal causal diamond in our cosmology, along some particular geodesic.  This matrix model description is the only non-hydrodynamic description of the so called "slow roll" era of inflationary cosmology in the HST model.  The free Dirac equation would preserve these constraints forevever.  

The dynamics of our model is causal, following what we called modular flow.  In order to incorporate the fast scrambling behavior of horizons, we have to add an interaction
\begin{equation} L_P^2 t^{-3} {\rm Tr} M^2 , \end{equation} where $M_i^j = \psi_i^A \psi^{\dagger\ j}_A $ is the square 
$t \times t $ matrix bilinear.  Under rotations it transforms as a sum of $p$ forms on the sphere and the trace acts as the integral, projecting out the $2$ form combinations.  In\cite{hilbertbundles} we argued that in order to preserve approximate two dimensional conformal invariance we should take the $M's$ to be currents on the stretched horizon and only use the product of the $0$ form and the $2$ form on the two sphere, which commute with each other.  

It is also important that the initial conditions be arranged such that the constraints on the large "final" diamond are distributed correctly among the nested diamonds that make it up.  This is just the statement that a particular localized object enters into the causal diamond of a particular geodesic at a particular proper time.  It's really an identification in terms of the initial data, of what geodesic in the background space-time we are talking about.  This peculiar point of view reflects the novel philosophy of our approach.  We are given a background space-time and we try to build a quantum mechanics whose hydrodynamics fits that space-time.  But when the fluctuations grow large enough to affect the geometry, as they have in our universe, we must go back and identify where each fluctuation occurred.  Our basic formalism is, apart from the cutoff on angular momentum, invariant under area preserving maps of the sphere, so it is these identifications of where various localized excitations are that determine the actual geometry.  

We assume that during the slow roll era the coarse grained hydrodynamics of our model is described by a flat FRW metric with scale factor $a(t)$ and slow roll factor $\epsilon (t) = - \frac{\dot{H}}{H^2} < 1 $ , for all time.  In order for this to be consistent with the assumption that the individual inflationary horizon volumes remain approximately non-interacting quantum systems throughout this era, the horizon volume must expand rapidly enough that the growth in the number of frozen degrees of freedom exceeds the fast scrambling rate.  This gives a constraint
\begin{equation} \epsilon > C [{\rm ln} ( M_P/H(t))]^{-1} . \end{equation}  The constant $C$ depends on the precise form of the fast scrambling Hamiltonian and is hard to calculate.  We will assume it is $o(1)$.  

We are of course not privy to the quantum initial conditions for the universe and can at best imagine a most probable distribution for them.  Our assumption of a coarse grained
FRW metric means that we must postulate a homogeneous isotropic distribution of horizon volumes of size $R_I$ on the hydrodynamic background.  We'll see in a minute the consequence of giving up this assumption.  

The dynamics inside each horizon volume is described by modular flow.  Since $R_I \gg L_P$ we can think of this semi-classically.  For wavelengths that are of order the horizon size, modular flow\cite{JV} in the maximal dS causal diamond is the same as static time translation.  Now lets define an action on the probability distribution of fluctuations of $\frac{\delta H}{H}$ and $\gamma_{ij}$ of the metric in co-moving gauge, assuming that those fluctuations come from fluctuations of the systems in the individual horizon volumes.  If we define and action that's the sum of the local flat coordinate time translations at the location of each horizon volume\cite{holoinfrev} then it commutes with the Euclidean group on the spatial sections of the FRW, and generates dS rescalings of coordinates locally around each point where there's a horizon volume.  We will be taking the spacing between those points at the initial FRW time\footnote{Note that this is NOT the initial time on the past horizon of any maximal causal diamond, where the quantum initial conditions are set.}  to be of order a few times $R_I$ (we'll be more precise in a moment).  Thus we have approximate invariance of the probability distribution under a group that is isomorphic to the subgroup of $SO(1,4)$ whose generators are $J_{04}, J_{4+0,i}, J_{ij}$.  This is enough to prove full dS invariance of the two point functions.  

The period after the slow roll era can be studied using a combination of classical cosmological perturbation theory and Hawking's analysis of black hole decay. Our matrix model description of the slow roll era was that along each of a set of geodesics of our cosmology, the universe began in the maximal $p= \rho$ state, but then the causal diamond of the geodesic expanded, leaving a system with entropy $\pi (R_I/L_P)^2 $ decoupled from the new degrees of freedom that were added, because the state of the system was constrained.  If we look on a global equal time proper time surface for all of those geodesics, those localized systems behave like black holes, according to the Carlip-Solodukhin ansatz.  The fact that $\epsilon$ is small implies that the distance between those black holes is not much larger than their radii.  After the slow roll era ends, the bulk description of the system is a dense fluid composed of black holes.  If the black hole masses are not close to equal there will be rapid mergers and the formation of large long lived black holes whose decay products cannot give rise to a conventional radiation dominated universe.  From the point of view of an underlying Hilbert bundle model there is nothing fined tuned about trying to model a homogeneous isotropic universe.  The initial conditions required are simply that in each fiber Hilbert space, at the same proper time, $R_I$, we have the same set of constraints, and the same expansion rate $a(t)$ during the brief slow roll period.  We can even allow $L_P/R_I$ differences in the number of constraints, so that the $S_N$ permutation symmetry between the diagonal blocks of the bilinear matrix variables is only approximate.  From the bulk point of view, this symmetry manifests as approximate homogeneity and isotropy of the black hole probability distribution in FRW coordinates.  Because the individual black holes are finite quantum systems, with Carlip-Solodukhin fluctuations in their entropy, we get a dS invariant prediction for the two point function of $\frac{\delta H}{H}$ on co-moving coordinate surfaces.  The gauge invariant scalar fluctuation $\zeta$ is related to this by a factor of $\epsilon^{-1}(t) $, which is evaluated at the time each scale crosses the horizon.  The tensor spectrum is exactly dS invariant and its normalization is fixed relative to that of $\frac{\delta H}{H}$ by using the Kerr black hole entropy formula to calculate the mean square angular momentum fluctuation per black hole, and compare this to the means square angular momentum fluctuation fluctuation over a horizon volume at the end of slow roll using the tensor fluctuation formula.  

In this model, scales smaller than the inflationary horizon do not make any sense in the power spectrum.  All fluctuations near that scale have to do with fluctuations of individual black hole properties and have no hydrodynamic interpretation, except for the Carlip Solodukhin prediction that they are associated with rescaling of the horizon coordinates.  Combining the local scale transformations for all black holes we get a translation invariant generator that rescales FRW spatial coordinates like the $J_{04}$ generator of the dS group.  This is the symmetry that acts on the probability distribution of black holes at wavelengths longer than $R_I$.  A fit to the CMB data has been obtained using this model\cite{tbsacmb} despite its many conceptual differences from field theoretic inflation.  

If we had not assumed approximate homogeneity up to inevitable fluctuations of finite quantum systems, the black holes would have quickly agglomerated into larger black holes with long lifetimes.  Most of their decay products would have been very low energy and nothing like the Hot Big Bang, nucleosynthesis, galaxy formation {\it etc.} would have occurred.  Thus we claim that, among holographic cosmologies that produce localized objects, those which produce a maximally uniform distribution of rather small black holes, which decay to a Hot Big Bang before too much large structure can form, are the only ones that produce a universe with galaxies in it.  

I will mention only briefly a few significant differences between the holographic inflation model and field theoretic inflation
\begin{itemize}
\item There is an early era of structure formation, before decay of the {\it inflationary black holes} sets off the Hot Big Bang.  If we postulate a discrete $Z_N$ gauge symmetry, whose lowest charge is carried by a black hole remnant with mass of order $M_P$, then these remnants can account for the dark matter.  It's quite plausible that a number of horizon size black holes can form by the time the radiation dominated era begins.  These will grow with the horizon, following the Novikov-Zeldovich scaling solution until the Carr-Hawking instability sets in. We conjectured that they might form the seeds for the supermassive black hole cores of the earliest galaxies.  It's marginally possible that the black hole decays can produce the required baryon asymmetry of the universe.  

\item The black hole decays produce gravitons that are a competing contribution to the "primordial" tensor fluctuations. Their power spectrum shares the shape of the scalar fluctuations including the $\epsilon^{-2}$ overall enhancement, but is suppressed by $1/g$ the number of effectively massless particle degrees of freedom at the reheat temperature, which is $\sim 10^{10}$ GeV.  We do not of course know what $g$ is at the present time.  If tensor fluctuations are detected we'll have to compare the data to both predictions from HST models.  It's unlikely that $g$ is finely tuned so that both are of the same order of magnitude.  

\item The three point fluctuations of the scalars are of the order expected from general theorems, but the tensor fluctuations are not the same as in field theory models.  Both "parity"-even terms are expected to be of the same order of magnitude.  The parity odd term might or might not vanish.  It's not clear whether this reflection symmetry is a property of the holographic models, even approximately.  At any rate, all non-Gaussian fluctuations are predicted to be small, as in conventional models.  Thus it's unlikely these differences will be detected in the near future.  

\item The conceptual explanation for homogeneity and isotropy is completely different in HST models than it is in field theory models, and does not require us to think about field theory modes whose wavelength was originally many orders of magnitude smaller than the inflationary horizon radius.  Homogeneity and isotropy basically come from the requirement that the universe contain localized objects in it, which are not large meta-stable black holes.  

\end{itemize}

\section{A Sobering Conclusion}

In this mini-review we've outlined the beginnings of a general hydrodynamic theory of quantum gravity.  It is based on three fundamental papers from the 1990s\cite{ted95}\cite{carlip}\cite{solo}, and a generalization\cite{BZ} of the latter two to arbitrary causal diamonds. They imply a universal pair of laws for the expectation value and fluctuation of the modular Hamiltonian $K$ of an arbitrary causal diamond in a spacetime that is a classical solution of Einstein's equations
\begin{equation} (\delta K)^2 = \langle K \rangle = \frac{A_{\diamond}}{4G_N} . \end{equation}
Here $ A_{\diamond}$ is the $d-2$ volume of the holographic screen of the diamond.  Carlip and Solodukhin and\cite{BZ} viewed this as coming from a $1 + 1$ dimensional CFT living on an interval (the stretched horizon) in the vicinity of the holoscreen.  The central charge of the CFT was $c \propto \frac{A}{4G_N}$ and there was a UV cutoff in the vicinity of the saddle point in the thermal integral over the Cardy density of states for large central charge\cite{BZ}.  

In a series of previous papers\cite{hst}, Fischler and the present author had proposed a description of the fluctuations of the geometry of the holographic screens of diamonds around their background values in terms of a fluctuating Dirac operator, using Connes\cite{connes} demonstration that Riemannian geometry is completely captured by the Dirac operator.  The idea was that the finite entropy would be imposed by a UV eigenvalue cutoff in the expansion of the fluctuating Dirac field in terms of background Dirac eigenspinors.  This can be put into the CS framework by making the expansion coefficients into massless Dirac fields on the stretched horizon.  A particular set of Thirring interactions makes the system into a fast scrambler.  

This general framework fits qualitatively with everything we know about quantum gravity.  It provides an explanation for the universal emergence of SUSY in asymptotically flat limits, a novel description of inflationary cosmology with new predictions for dark matter, baryogenesis and perhaps even early galaxy formation.  It resolves the firewall paradox and begins to give us a hint at how QFT emerges as an approximation to QG in appropriate circumstances.  The biggest failure of the formalism is a clear mechanism for enforcing the QPR, which would guarantee that symmetries emerged in appropriate limits.  

A major feature of the formalism, which distinguishes it from a lot of rhetoric about a hypothetical universal theory of quantum gravity is that it is manifestly {\bf Background Dependent}.  It's been my opinion, since the discovery of non-perturbative $1 + 1$ dimensional string models in the early 1990s, and particularly the AdS/CFT correspondence, that the idea of background independence in quantum gravity was a mistaken one.  Jacobson's identification of Einstein's equations as the hydrodynamics of the area law makes that even more clear.  We do not think that all systems that exhibit Navier-Stokes behavior have the same microscopic dynamics.  A particular feature of background dependence, which has not come into play in this review has to do with the local physics of constrained states.  The dynamics on individual holographic screens is invariant under a fuzzy version of volume preserving maps of the screen. However, the embedding of nested screens into the background geometry tells us the "angles" at which the constraints live on the screens, and thus determine the trajectories of jets of particles through space-times.  The quantum in quantum gravity, has to do with fluctuations around a particular hydrodynamic background, which determines the empty diamond density matrices and the geometric properties of the constrained states we call particles. If the detailed conjectures in this paper are right, the background determines the nature of the quantum variables, and much about their interactions.  

Another issue on which I think the hydrodynamic point of view throws light is the role of non-trivial topologies in the effective Euclidean gravitational path integral. Swingle and Winer\cite{sw} showed that the "ramp" in the SYK model, which is reproduced by a non-trivial topology in JT gravity, and more generally the ramp in any chaotic system's spectral form factor, is reproduced by the hydrodynamic approximation to the system.  The "paradoxical" failure of factorization is explained by the fact that hydrodynamics calculates time averaged quantities.  A folk-theorem in quantum chaos relates time averaging to averaging over Hamiltonians.  This can be made a little more explicit using the fact that the fluctuating hydrodynamics of a quite general quantum system can be written as a Euclidean path integral for the energy density of a large region with a standard second order kinetic term because the energy density of a large enough region is a "slow mode" of the dynamics.  The action also contains a term proportional to the local entropy $S(E)$.  Using a trick introduced by Verlinde\cite{hv}, one can write the connected spectral correlators as integrals over fields that live on a hyperbolic disk with multiple punctures.  The result looks like a double scaled random matrix integral, expanded to leading order in large $N$. The spectral density is determined by the original chaotic system\cite{tbhydrowormhole}.  

However, the formalism reveals a rather disturbing feature of any attempt to combine quantum mechanics with gravitation.  Science is supposed to be about comparing theory with experiment.  This has come to mean comparing precise mathematical predictions with measurements that could at least in principle be made as precise as one wanted.  In modern language, von Neumann's criterion for a good detector in the context of quantum mechanics was that it should have one "semi-classical q-bit" for every q-bit in the system it was supposed to measure.  A semi-classical q-bit is a large subsystem of a large quantum system, which approximately obeys the laws of classical statistical mechanics.  More precisely, there's some large integer $N$ such that interference effects in the quantum mechanics of the semi-classical q-bits are of order $e^{-N}$ or smaller, whereas fluctuations of the q-bits are of order $N^{-p}$ with $p$ of order $1$.  Typically $N > 10^{20}$.  In the real world we know of two types of large quantum systems, black holes, and those that are described as complex bound states in quantum field theory.  Of these, only the latter has many subsystems that can serve as semi-classical q-bits.
The field theoretic states inside a causal diamond are vastly outnumbered by the boundary states that make up most of the entropy.  Thus, there is, {\it in principle} no way to experimentally test the complete theory of the causal diamond proposed here, much less one that attributes an infinite dimensional operator algebra to the diamond.  

Tests of these ideas will thus be mathematical in nature.  In particular the biggest challenge is to find a way to guarantee the implementation of the QPR, because this would imply that in the case of asymptotically flat and AdS space-times, the appropriate asymptotic symmetries are realized.  For the AdS case this can in principle be done via the TNRG, though carrying out the actual computations to find the TNRG transition maps is probably impossible in practice for any realistic model.  For asymptotically flat space we know how to take a limit of the operator algebra to make a (generalized) Lorentz covariant Awada-Gibbons-Shaw super-BMS algebra\cite{tbags}, but proving the representation space of this algebra is a separable Hilbert space and the transition operator between past and future is Lorentz covariant is still a formidable task.  

In the case of space-times that are asymptotically dS in the future, the mathematical framework is much more ambiguous.  What is certain is that there is no Hilbert space on which any of the elements of the dS isometry group act.  In the Hilbert bundle formulation, most of these map one fiber of the bundle into another.  Even those isometries that preserve a given geodesic do not correspond to conserved quantities. Consider in dimension $d \geq 3$ a massive quantum particle whose wave function is initially concentrated around a particular geodesic.  In a time of order $R {\rm ln} (m R)$ it spreads uniformly over the horizon.  Neither energy, charge, nor angular momentum is conserved.  
For $ d \geq 4$\footnote{Since all known models with vanishing c.c. are exactly supersymmetric and there are no known ways to find de Sitter solutions of supergravity in $d > 4$, it is very likely that the only consistent quantum gravity models of dS space have $d \leq 4$.  Indeed, we're not even sure that models with $d = 3$ exist. }, one can have longer lived localized excitations in dS space because of the existence of meta-stable gravitational bound states, colloquially known as local groups of galaxies.  These do not travel on geodesics and their motion is subject to the vagaries of cosmic history.  However the trajectory of the center of mass of such a group is a semi-classical variable whose motion is rendered decoherent by entanglement with the myriad subsystems that make up the local group.  It defines a universal proper time coordinate and the other semi-classical trajectories provide a plethora of detectors for many quantum states in the causal diamond of the group.  That causal diamond is defined roughly by the beginning of the universe (since much data about that beginning can be gleaned from the Cosmic Microwave Background, Large Scale Cosmological Structure {\it etc.}) and ends at the point where the local group collapses into a black hole.   At that point it continues to exist, but its utility as a detector has come to an end because it has very few semi-classical subsystems that can reliably store and record information.  Thus, no theory of mathematical dS space is of much  use for telling us much about our own universe, and much of what we will be able to learn from a theory of the cosmos will at best tell us how probable the particular events we see in our own local group are, among all the possible local groups of galaxies in the (temporarily) visible universe.

\vskip.3in
\begin{center}
{\bf Acknowledgments }
\end{center}
 The work of T.B. was supported by the Department of Energy under grant DE-SC0010008. Rutgers Project 833012.

%%%%%%%%%%%%%%%%%%%%%%%%%%%%%%%%%%%%%%%%%%%%%%%%%%%%%%%%%%%%%%%%%%%%%%%%%%%%%%%%%%%%%%%%%%%%%%%%%%%%%%%%%%%%%%%%%%%%%%%%%%%%%%%%%%%%%%%%%%%%%%%%%%%%

% \bibliographystyle{utphys}
% \bibliography{fuzzy_refs}

%%%%%%%%%%%%%%%%%%%%%%%%%%%%%%%%%%%%%%%%%%%%%%%%%%%%%%%%%%%%%%%%%%%%%%%%%%%%%%%%%%%%%%%%%%%%%%%%%%%%%%%%%%%%%%%%%%%%%%%%%%%%%%%%%%%%%%%%%%%%%%%%%%%%
%%%%%%%%%%%%%%%%%%%%%%%%%%%%%%%%%%%%%%%%%%%%%%%%%%%%%%%%%%%%%%%%%%%%%%%%%%%%%%%%%%%%%%%%%%%%%%%%%%%%%%%%%%%%%%%%%%%%%%%%%%%%%%%%%%%%%%%%%%%%%%%%%%%%

\end{document}